\documentstyle[aaspp4]{article}

\begin{document}

\title{TIMING IN THE TIME DOMAIN: CYGNUS X-1 \footnote{To appear in {\sl CJAA
} \bf1,
313 (2001)}}
\author{Ti-pei Li$^{1,2}$}  
\affil{ $^{1}$ Department of Physics \& Center for Astrophysics, Tsinghua University, Beijing\\
$^{2}$ High Energy Astrophysics Lab., Institute of High Energy Physics, \\
Chinese Academy of Sciences, Beijing}

\begin{abstract}
Quantities characterizing temporal property, e.g. power density, coherence,
and time lag,  can be defined and calculated directly in the time domain
without using the Fourier transformation. Spectral hardness,  
variability duration, and correlation between
different characteristic quantities on different time scale can be studied 
in the time domain as well.
The temporal analysis technique in the time domain is 
a powerful tool, particularly in studying rapid variability on short time scales 
(or in high frequencies). Results of studying variabilities of X-rays from Cyg X-1
with the analysis technique in the time domain and {\sl RXTE} data reveal valuable clues 
to understanding production and propagation processes of X-rays and structure of 
accretion disk in the black hole system. 
\end{abstract}

\keywords{methods: data analysis  --- stars: individual 
(Cygnus X-1) --- X-rays: stars}

\section{INTRODUCTION}
Studying short time scale variability of the X-ray emission of black-hole systems
and low-mass X-ray binaries is an important approach to understanding the emitting
region and emission mechanism of high-energy photons.
The time-averaged spectra of hard X-rays from Cyg X-1 and other black-hole candidates 
are relatively well
explained in terms of a simple model: hard X-rays result from the Comptonization
of soft photons in a hot electron cloud of constant temperature and optical
depth (e.g. \cite{sha76,sun80}).
It is difficult to learn more about the process 
of high-energy emission in these  objects if we have only the spectral information.
Complex rapid fluctuation of the X-ray emission is a common characteristic 
of black-hole systems and low-mass X-ray binaries (\cite{kli95}).   
Some properties of aperiodic variability of X-rays from Cyg X-1 revealed by timing analysis, 
e.g. time lags (\cite{miy92,cui97,cra98,now99}) 
and coherence (\cite{vau97,cui97}) between different energy bands, 
are difficult to interpret by the simple 
Comptonization model and provide strong constraints on theoretical models, 
which are attracting an increasing amount of attention. 

The Fourier spectrum technique is most widely used in timing analysis.
Let $x(t_k)$ be the photon counts during a time interval $(t_k,~ t_k+\Delta t)$, 
where $t_k=k \Delta t$,~ $k=0,1,...,N-1$, and  $T=N\Delta t$ the observation duration.
 The discrete Fourier transform of the time series (light curve) $x(t_k)$ is
\begin{equation} 
X(f_j)=\sum_k x(t_k) e^{-i2\pi f_jk\Delta t}, \hspace{5mm} f_j=j/T 
\end{equation}
The Fourier power spectrum is usually used to describe the variation
amplitude at different frequency $f_j$
\begin{equation} P_j=|X(f_j)|^2 \end{equation} 
From two counting series, $x_1(t_k)$ and $x_2(t_k)$, observed simultaneously in two energy 
bands at times $t_k$, and their Fourier transforms $X_1(f_j)$ and $X_2(f_j)$, one can 
construct the cross spectrum $C(f_j)=X_1^*(f_j)X_2(f_j)$, with argument the phase 
difference between  the two processes at frequency $f_j$, or the time lag
of photons in band 2 relative to that in band 1 
\begin{equation} \tau (f_j)=\arg [C(f_j)]/2\pi f_j \end{equation} 
Dividing the light curve into several segments and calculating their Fourier spectra 
and cross spectra for each segment, the coherence coefficient is then defined as 
\begin{equation} r(f)=\frac{|<C(f)>|}{\sqrt{<|X_1(f)|^2><|X_2(f)|^2>}} \end{equation}
where angle brackets denote an average over the segments. The coherence coefficient 
is used as a
measure of the degree of linear correlation between the two time series at
a Fourier frequency $f$. If the ratio of two Fourier transforms at a frequency $f$,
$H(f)=X_2(f)/X_1(f)$, is the same for all segments of the two  processes,
or, equivalently
\begin{equation} x_2(t)=\int h(t-\tau)x_1(\tau)d\tau \end{equation}
then 
\begin{equation} r(f)=1 \end{equation}
the processes are said to be coherent at frequency $f$.

Quite a number of important results in the study of spectrum structure, hard X-ray lags and 
coherence between high and low energy bands of X-rays from Cyg X-1 and other objects
 have been obtained 
with the above Fourier techniques. However, Fourier analysis can not replace 
studying variability directly in the time domain. Except periodic and quasi-periodic 
processes, their is no direct correspondence between a structure in the Fourier spectrum 
and the physical process on a certain time scale. The power density, time lag or coherence 
at a given Fourier frequency can result from contributions by different processes on different time scales.
In addition, the Fourier transformation is sensitive to dead times and data gaps
caused by various reasons; and the window effect limits the performance of Fourier technique  
in studying rapid variability in high frequency region.          

Without using the Fourier transformation, we can also calculate quantities characterizing  
temporal property, e.g.,  power density, time lag and 
coherence between different energy bands, and, furthermore, we can study the hardness ratio, variability duration and
correlation between different quantities on different time scales  directly in the time
domain. In Section 2 various characteristic quantities are defined and applied to analyzing
light curves of Cyg X-1 observed by ${\sl RXTE}$. In Section 3 coherence and spectral hardness
of shot component on different time scales are studied.  
Our results show that 
with the aid of timing analysis in the time domain important characteristics of physical
processes on different time scales can be revealed, and that the timing technique in the
time domain is particularly powerful in studying variability behavior on short time scales (or
in high frequencies). Relevant discussions are made in Section 4.     

\section{CHARACTERISTIC QUANTITIES IN THE TIME DOMAIN}
In this section we define some quantities characterizing temporal property of
emission on a given time scale in the time domain, including the power density,   
 time lag, variability duration, coherence, and explain how to apply 
them to the study of rapid variability through analyzing light curves of Cyg X-1 in different
emission states observed by ${\sl RXTE}$. For a given time scale $\Delta t$, we
produce a counting series (light curve) $x(k)$ with time step $\Delta t$.
The time series $x(k)$ is divided into $M$ segments, each segment includes 
$N$ successive counts. From the $N$ counts $x(i_0+1), x(i_0+2), ..., x(i_0+N)$ of 
segment $i$, we calculate the quantity under study, 
\begin{equation} y(i)=f(x(i_0+1), x(i_0+2), ..., x(i_0+N)) \hspace{5mm}i_0=(i-1)N
\end{equation}
After obtaining $M$ values of $y$ from $M$ data segments, the average $\bar{y}$ and
its standard deviation $\sigma (\bar{y})$ can be derived:
\begin{equation}  \bar{y}=\frac{1}{M}\sum_{i=1}^{M} y(i) \hspace{8mm} 
\sigma (\bar{y})=\sqrt{\sum_{i=1}^{M}(y(i)-\bar{y})^2/M(M-1)}
 \end{equation}
 Usually we can use some convenient statistical methods based on
the normal distribution to make statistical inference, e.g. significance test,
on $\bar{y}$. For the case of short time scale  $\Delta t$, although the number  $x$
of counts per bin may be too small for it to be assumed as a normal variable, it is easy
from a certain observation period to get the total number $M$ of segments large enough 
to satisfy the condition for applying the central limit theorem in statistics    
and using the normal statistics for the mean $\bar{y}$.

{\bf (1) Power Density}

Let $x(k)$ is a counting series obtained from a time history of observed photons
with a time step $\Delta t$, $r(k)$ is the corresponding counting rate series,
the variation power in the light curve $x(k)$ is
\begin{equation}
P(\Delta t)=\frac{\mbox{Var}(x)}{(\Delta t)^2}=\frac{\frac{1}{N}\sum_{k=1}^{N}(x(k)-\bar{x})^2}
{(\Delta t)^2}=\frac{1}{N}\sum_{k=1}^{N}(r(k)-\bar{r})^2
\hspace{7mm} \mbox{rms$^2$} 
\end{equation}
where $\bar{x}=\sum_{k=1}^{N}x(k)/N, \bar{r}=\sum_{k=1}^{N}r(k)/N$.
The power $P(\Delta t)$ of a certain series of photon events is a function of the time step 
$\Delta t$. It is obvious that the variation on a time scale 
small than $\Delta t$ makes no contribution to the power $P(\Delta t)$, but variation
on a time scale greater than or equal to $\Delta t$ does. The power density $p(\Delta t)$
in the time domain can be defined as the rate of change of $P(\Delta t)$ with respect to the
time step $\Delta t$. From two powers, $P(\Delta t_1)$ and $P(\Delta t_2)$,
at two time scales, $\Delta t_1$ and $\Delta t_2$ and $\Delta t_2>\Delta t_1$, we can evaluate the power density at
$\Delta t=(\Delta t_1+\Delta t_2)/2$ approximately by
\begin{equation}
p(\Delta t)=\frac{P(\Delta t_1)-P(\Delta t_2)}{\Delta t_2-\Delta t_1}
 \hspace{7mm} \mbox{rms$^2$/s} \end{equation}
For a noise series where the $x(k)$ follow the Poisson distribution the noise power
\begin{equation}
P_{noise}(\Delta t)=\frac{\mbox{Var}(x)}{(\Delta t)^2}=\frac{<x>}{(\Delta t)^2}
=\frac{r}{\Delta t} \hspace{7mm} \mbox{rms$^2$} 
\end{equation}
where $r$ is the expectation value of counting rate which can be estimated by 
the global average of counting rate of the studied observation.
The noise power density  at $\Delta t=(\Delta t_1+\Delta t_2)/2$
\begin{equation}
p_{noise}(\Delta t)=\frac{P_{noise}(\Delta t_1)-P_{noise}(\Delta t_2)}
{\Delta t_2-\Delta t_1}
 =\frac{r}{\Delta t_1 \Delta t_2} \hspace{7mm} \mbox{rms$^2$/s} 
\end{equation}
The signal power density can be defined as
\begin{equation}
p_{signal}(\Delta t)=p(\Delta t)-p_{noise}(\Delta t) \hspace{7mm} \mbox{rms$^2$/s} 
 \end{equation}
and the fractional signal power density
\begin{equation}
p'_{signal}(\Delta t)=\frac{p_{signal}(\Delta t)}{r^2} \hspace{10mm} \mbox{(rms/mean)$^2$/s}
 \end{equation}

To study the signal power density in the time domain over a background 
of noise in an observed photon series we divide the observation 
into $M$ segments. For each segment $i$ the signal power density $p_{i,signal}(\Delta t)$ is
 calculated by Eq.~(14) and then the average power density of the studied observation
$\bar{p}_{s}=\sum_{i=1}^M p_{i,signal}/M$ and its standard deviation 
$\sigma(\bar{p}_{s})=\sqrt{\sum_{i=1}^{M}(p_{i,signal}-\bar{p}_{s})^2/(M(M-1))}$.

Two kinds of signal are used to compare the powers so defined with the Fourier spectrum.
One is a triangular signal $s(t)$ with a period of 5~s and peak rate of 1000~cts/s. From $s(t)$ 
it is easy to make the light curve of pure signal with any time step,
as shown in the top panel of Fig.~1 for a piece of the signal light curve with $\Delta t=0.01$~s, 
and to calculate the expected signal power density spectrum in the time domain, shown by the solid line in the
bottom panel of Fig. 1. A simulated 1000~s light curve $x(k)$ with time step 1~ms is made
by a random sampling of the theoretical light curve of signal with Poisson fluctuation plus a Poisson 
noise with mean rate 5000~cts/s.  The middle panel of Fig.~1 shows a piece of
the light curve with $\Delta t=0.01$~s obtained  from the simulated 1~ms light curve.
For each segment of the simulated light curve we calculate the total power density $p(\Delta t)$ 
by Eqs. (9), (10) and the noise power density $p_{noise}(\Delta t)$ by Eq. (12),
then get the signal power density  $p_{signal}(\Delta t)=p(\Delta t)-p_{noise}(\Delta t)$.
In the bottom panel of Fig.~1 the plus signs mark the average signal power densities 
at different time scales.
For the same light curve of 1 ms time bin we also calculate the Leahy density $w(f_j)=2|X_j|^2/X_0$
 for each $T=4.096$ s segment where $X_j$ is the Fourier amplitude at frequency $f_j=j/T$ determined 
from a 4096-point FFT. It is well known that the noise Leahy density $w_{noise}=2$,
 so the signal Leahy density $w_{signal}(f_j)=w(f_j)-2$ and the Fourier signal power density 
$p_{F,signal}(f_j)=w_{signal}(f_j)X_0/T$. The dashed line in the bottom panel of Fig. 1 
shows the average Fourier signal power density as a function of timescale $\Delta t_j=1/f_j$ 
obtained with the transformation $p(\Delta t_j)=p_F(f_j)f_j^2$.
 From Fig.~1 we can see that 
 the signal power densities determined by Eqs.~(9)-(13) can reflect  the real power distribution
in the triangular signal in the time domain but that those by the Fourier analysis can not.
The structures in the representation of Fourier power spectrum in the time domain, 
e.g.,  the peak at the short time scale region
of $\Delta t<0.03$~s, are needed by the mathematical construction of  the light curve
with sinusoidal functions, but they do not represent  the real signal powers in processes occurring in the time domain. 

The other kind of signal examined is a stochastic process. We use an autoregressive (AR)
 process  of first order 
$u(k)=a\cdot u(k-1)+\epsilon$ to make the signal light curve $s(k)$ with time bin
 $\Delta t=0.01$~s, $s(k)=c\cdot u(k) + r_s\Delta t$, shown in the top panel of Fig.~2,
where the relaxation time $\tau = -\Delta t/\log | a | = 0.1$~s,
$\epsilon$ is a Gaussian random  variable with mean 0 and variance 1, 
the mean rate of signal
$r_s=2000$~cts/s and $c=4.8$.  The final observed light curve $x(k)=s(k)+n(k)$
with $n(k)$ a Poisson noise with mean rate 5000~cts/s,
shown in the middle panel of Fig.~2. To partly eliminate the contribution of the system noise 
$\epsilon$ we use $s'(k)=c\cdot u(k-1) + r_s \Delta t$ instead of $s(k)$ to calculate the 
intrinsic power density of signal. In the bottom panel of Fig.~2 the solid line
shows the expected variation power density distribution of the signal $s'(k)$,
the plus signs indicate excess power densities in the light curve $x(k)$ estimated by Eqs.~(9)-(13)
 and the dashed line by FFT. Figure 2 shows that the proposed procedures in the time domain
are able to extract variation powers in a stochastic process from noisy data
and obtain an excess power spectrum   
more exactly than by using the Fourier spectral representation. The Fourier spectrum
significantly underestimate the signal powers in the region of short time scales (high frequency region) 
and overestimate those in the region of long time scales.

Now we make a power spectral analysis in the time domain for different 
states of Cyg X-1 X-ray emission.
On 1996 May 10 (day 131 of 1996) the All-Sky Monitor on the Rossi X-ray Timing 
Explorer ($RXTE$) revealed 
that Cyg X-1 started a transition from the normal hard state to a soft 
state. After reaching the soft state, it stayed there for about 2 months 
before going back down to the hard state (\cite{cui97}). 
During this period 11 pointing observations of Cyg X-1 were made by
$RXTE$.  We use one observational run of PCA detector on-board $RXTE$ for
each of the four states of Cyg X-1, hard-to-soft transition, soft, soft-to-hard
transition and hard. The total duration of each run is about 2000 s.
For 18 time steps between 0.001 s and 2.5 s,  we  make the corresponding
light curves in the 2-13 keV band, and pick out all the ineffective data points 
caused by failure in the satellite, detector or data accumulation system. 
The effective data of an observation is divided up into $M$ equal segments with
$N=100$ data points each.  If the segment number $M<100$ in the case of large time scale,
let $M=100$ and  decrease the number $N$ of data points in each segment accordingly.
For each segment of the light curve with time bin $\Delta t_1$ we calculate the powers
at two time scales $\Delta t_1$ and $\Delta t_2=2\Delta t_1$ by Eq.~(9)  
and the power density at $\Delta t=1.5\Delta t_1$ by Eq.~(10). The corresponding noise power density is
calculated by Eq.~(12) with $r$ set equal to the average observed counting rate.
The plus signs in Figure 3 show the distributions of the average signal power density of Cyg X-1 
calculated directly in the time domain for four $RXTE$ observations. All but one of the power densities 
of Cyg X-1 shown in Fig.~3 have statistical
significance $s=\bar{p}_{s}/\sigma(\bar{p}_{s})>10$.
The narrow panel under each plot of Cyg X-1 in Fig.~3 shows 
the corresponding results of a fake light curve of Poisson noise with mean the average rate $r$ of
 the observed Cyg X-1 light curve. For each observation and each time step $\Delta t$ studied 
1000 fake light curves
are produced and their signal power densities calculated. In a total of 72000 trials 
the number of events with significance $s\ge 2$ is 1561 (the expectation from the normal 
distributions is 1638 ), 107  with $s\ge 3$ (expectation 97.2), 2 with $s\ge 4$ (expectation 2.3) 
and no event has
$s\ge 5$. Therefore  $s=\bar{p}_s/\sigma(\bar{p}_{s})$ can be seen approximately 
as a standard normal variable and be used in statistical significance tests. 
The  corresponding Fourier power densities of each Cyg X-1 observations are also   
shown in Fig. 3 (dots).  The Fourier spectra are significantly lower than the power spectra of
time domain in the time scale region of $\Delta t<0.1$~s, which  is not a surprise  
as the rapid variability of Cyg X-1 can be 
described by an AR process with a relaxation time $\tau$ of 
about 0.1~s approximately (\cite{pot98}) and as we show above that 
the Fourier spectral representation significantly underestimate the powers of AR process 
in the region of time scale shorter than the relaxation time. 
Comparing the two kinds of power density spectra, the Fourier spectrum and the spectrum analyzed in
the time domain, may help reveal intrinsic nature of the radiation process under study.
The two kinds of spectra
have been derived for a sample of X-ray binaries. The results
show that the property that the Fourier spectra is significantly different 
from the power spectra of
time domain in the short time scale region is found in the black hole binaries but
not in the neutron star binaries. Figure 4 shows, as an example, for the neutron star binary 
4U0614+091 the power density spectra analyzed both in the time and frequency domain. 
The complete results will be reported in a separate paper.      

To show the shapes of power spectra more clearly we multiply the signal power densities by
the corresponding time scales and draw the distributions of $p_s(\Delta t)\cdot\Delta t$ for
 four different states of Cyg X-1
in Figure 5. As indicated in Fig.~5, the distribution shapes 
of power spectra of the soft and hard states are similar to each other and the absolute sizes
of power densities are greater for the soft state than for the hard state, although 
the fractional power densities of the soft state are smaller. The power densities
of the transition states are distributed closely to those of the soft state in the time scale region 
of $\Delta t<0.1$~s, but  the two diverge in the region of $\Delta t>0.1$~s

{\bf (2) Hardness}

For two light curves ${x_1}$ in the energy band 1 and ${x_2}$ in band 2
with the same time step $\Delta t$, and under  the condition that both $x_1(i)$ and $x_2(i)$ 
in the same bin $i$ are greater than zero, we can calculate a hardness ratio,
\begin{equation} h_i=\frac{x_2(i)}{x_1(i)} \end{equation}
The average hardness ratio and its standard deviation on the time scale $\Delta t$ 
can be derived from a series of $h_i$.
To check the utility of the above definition of hardness  we calculate $h(\Delta t)$
for two fake light curves of Poisson noise, which are created by the standard $RXTE$ ftools 
with mean intensity 6.7 cts/ms for channel 1 and 3.3 cts/ms for channel 2 respectively. 
The light curve length
is 2000 s in the case of $\Delta t \geq 0.05$ s and 200 s for shorter $\Delta t$.
The left panel of Figure 6 shows the distribution of average $<h>$ vs. $\Delta t$ for  the
fake  light curves. From this result one can see that Poisson noise can cause
the average hardness ratio to decrease in the short time scale region, and this ratio 
defined by Eq. (15) is not a suitable quantity for studying hardness of physical process on
different time scales. An alternative definition of hardness ratio is
\begin{equation} H_i=\frac{x_2(i)-x_1(i)}{x_2(i)+x_1(i)} \end{equation}
The right panel of Figure 6 shows the distribution of $<H>$ vs. $\Delta t$ from the 
fake  light curves, where the hardness ratio keeps constant,
indicating that the ratio $H$ defined by Eq. (16) is a proper quantity 
for studying spectral hardness on different time scales. 
 The statistical averages of the hardness ratio
$H_i$ defined by Eq. (16)  for four $RXTE$ observations of Cyg X-1 and different time scales are
calculated and shown by the filled circles in Figure 7.

{\bf (3) Coherence}

Let $\{x_1\}, \{x_2\}$ be two background-subtracted light curves 
with time step $\Delta t$ for the energy band 1 and 2, respectively.  
Divide each  of two light curves $\{x_1\}, \{x_2\}$
into several segments,
the corresponding segments in different energy bands have the same time interval.
For a segment $i$ the following coefficient may be used to measure the degree of
linear correlation, i.e. coherence, between the two bands  
\begin{equation} r_x(i)=\sum_j x_1(j)x_2(j)/\sqrt{\sum_jx_1^2(j)\sum_jx_2^2(j)}
\end{equation}
The summations in Eq. (17) are only taken over such bins in which
both counts are effective and greater than zero.  
After calculating the average  and its standard deviation for
each segment, the final result $r_x(\Delta t)=\bar{r}_x$ and its error 
for the observation concerned can be derived, where $\Delta t$ is the time step of light curves.
The left panel of Figure 8 plots the distribution of $r_x$ vs. $\Delta t$ from
the fake light curves. As the values of $r_x(\Delta t)$ from the fake light curves are significantly 
greater than zero in the time scale region considered, the quantity $r_x(\Delta t)$  
defined by Eq. (17) is not a proper one for describing the correlation property between
two light curves.  The difference sequences $d_1(j)=x_1(j+1)-x_1(j)$,  $d_2(j)=x_2(j+1)-x_2(j)$ 
may more suitable than the original light curves $x_1(j)$, $x_2(j)$ for studying correlation 
property of variabilities in two channels. For a group $i$ of differences the coherence 
coefficient can be defined as
\begin{equation} r_d(i)=\sum_j d_1(j)d_2(j)/\sqrt{\sum_jd_1^2(j)\sum_jd_2^2(j)}
\end{equation}
The right panel of Figure 8 plots the distribution of $r_d$ vs. $\Delta t$ from
the fake light curves. That all the values of  $r_d(\Delta t)$ from the fake light curves are
near zero indicates that the coherence coefficient $r_d(\Delta t)$ can be used to
measure the correlation of variabilities other than statistical fluctuation in 
the two time series
 
The filled circles in Figure 9 show the coherence $r_d(\Delta t)$, evaluated by Eq. (18) 
for X-rays between energy bands 2-6.5 keV 
and 6.5-13 keV (or 2-5 keV and 5-13 keV for the hard state) of Cyg X-1 as a function 
of the time scale. From Fig. 9 one can see that when time scale $\Delta t>0.1$ s 
the intensity variabilities in two energy bands are nearly in perfect coherence,
$r_d(\Delta t)\simeq 1$, for all states.
It should be noted that unity coherence in the time domain, $r_d(\Delta t)=1$, indicates  
a linear correlation between the variabilities of intensity in the two energy bands,
$d_2(t)=hd_1(t)$, where $h$ is a constant during  the observation,
which is a stronger constraint than Eq. (5)  from unity coherence, $r(f)=1$,
in the frequency domain.

{\bf (4) Time Lag}

Correlation analysis is a technique for studying relative time delays
between two energy bands. For two groups of data 
$x_1(i)$, $x_2(i)$, $i=1,...,N$, in the same time period the cross-correlation function of
the zero-mean time series   is usually defined as
\begin{equation} \mbox{CCF}(k)=\sum_i v_1(i)v_2(i+k)/\sigma(v_1)\sigma(v_2) 
\hspace{3mm} (k=1,\pm 1, ...) \end{equation}
where $v(i)=x(i)-\bar{x}$, $\sigma^2(v)=\sum_i[v(i)]^2$. The corresponding time lag of CCF($k$)
is $\tau=k\Delta t$, where $\Delta t$ is the width of a time bin.   
With the traditional CCF defined above, it is difficult to measure 
time lags $\tau\le \Delta t$.  
To get necessary resolution for time lags we modify the above 
definition of CCF so that we can use any value  $\delta t$ for the
time lag (\cite{lit99}). On a time scale $\Delta t$ the modified cross-correlation function MCCF$(\tau)$  
at time lag $\tau$ is defined as
\begin{equation} \mbox{MCCF}(\tau)=\sum_i v_1(i\Delta t)v_2(i\Delta t+\tau)/\sigma(v_1)\sigma(v_2) 
\end{equation}
where $v(t)=x(t)-\bar{x}$, $x(t)$ is the counts in the time interval
$(t,~t+\Delta t)$. Then, we can use a time lag step $\delta t$ smaller than 
the time scale under study $\Delta t$, $\delta t<\Delta t$, and evaluate 
the values of MCCF($\tau$) at $\tau=k\delta t$, $k=0,\pm1,...$. If the function
MCCF($k$)/MCCF(0) has maximum at $k=k_m$, the time delay of the energy band 2 
relative to the band 1 $\Lambda =k_m\delta t$.

To test the utility of the above MCCF technique of estimating time lags we produce two photon event
serieses of length 1000 s with a known time lag between them.
 The series 1 is a white noise series with average rate 1000 cts/s. The series 2 consists 
of the same events in series 1 but each event time is delayed 5 ms. 
Besides the signal photons mentioned above, the two serieses are given additional independent noise events 
at average rate 100 cts/s. We use Fourier cross spectrum with 1ms lightcurves and 4096-point FFT 
and MCCF in the time domain to estimate the time lags between the two serieses at time scales 
$\Delta t$ from 1 ms to 2 s (in Fourier analysis  we take Fourier frequency $f=1/\Delta t$) and show the results in figure 10. From Fig. 10  one can see that the Fourier analysis fails 
for the short timescale region ($\Delta t$ from 1 ms to 0.02 s or $f$ from 50 Hz to 1000 Hz) 
but the MCCF in the time domain is successful.  

We calculated time lags between the 2-6.5 keV and 6.5-13 keV energy bands 
(or 2-5 keV and 5-13 keV for the hard state) for different time step light curves of Cyg X-1 
in different states observed by PCA/$RXTE$. The search for the maximum of MCCF 
was performed in the region of time lag $\tau$ within $0\pm1.5\Delta t$ for a given time scale 
$\Delta t$. The obtained time lag distributions of Cyg X-1 in its different  
states are shown in Figure 11. 
In the calculation one run of observation was divided into several groups. 
For short time scale the fluctuation of counts in a bin 
is large and a greater number of data points $N$ per group is needed to obtain 
a reliable value of MCCF. 
For this purpose we let each group to satisfy the condition of  $N\Delta t>10$ s.

{\bf (5) Variability Duration}

On a time scale $\Delta t$ the autocorrelation function of a light curve 
in an energy band $l$ at a time lag $\tau$ can be defined as
\begin{equation} \mbox{MACF}_l(\tau)=\sum_i v_l(i\Delta t)v_l(i\Delta t+\tau)/\sigma^2(v_l) 
\end{equation}
The FWHM of MACF$_l(\tau)$, $W_l$, can be taken as a measure of the variability duration 
of the light curve. 
Figure 12 shows MACF widths of Cyg X-1 in different 
states on different time scales, with pluses for the relative MACF widths 
for the low-energy band,
$W_1/\Delta t$, and triangles for the high-energy band, $W_2/\Delta t$. Figure 13 
shows the distributions of the ratio $W_2/W_1$ of MACF widths in the two energy bands.

\section{HARDNESS AND COHERENCE OF SHOT COMPONENT}
  
 It is interesting to see a time scale dependence of the hardness of the shot component
of Cyg X-1. We take the following procedure to distinguish the shot component of a light curve.
Divide a light curve of an observation into several groups  
of length $20\Delta t$ each, where $\Delta t$ is the time step of the light curve. 
For a group with $N_1$ effective 
counts $x(t_1),~x(t_2),~ ..., ~x(t_{N_1})$, $N_1\leq 20$, produce another data group 
$x'(t_1)=x(t_1),~ x'(t_2)=x(t_2),~ ..., ~x'(t_{N_1})=x(t_{N_1})$, then:
(1) fit $\{x'\}$ to a quadratic polynomial $f(t)$;
(2) find the maximum point $t_m$ of $x'(t)-f(t)$, let $x'(t_m)=f(t_m)$;
(3) calculate $\chi^2=\sum_i(x'(t_i)-f(t_i))^2/f(t_i)$, if $\chi^2>(N_1-1)$
then repeat the above procedure starting from (1), until the condition of
$\chi^2\leq (N_1-1)$ is satisfied. Let the shot component $x_s(t_i)=x(t_i)-x'(t_i)$, $i=1,2,...N_1$.  
The solid curve of the left panel of
Fig. 14 shows a group of 2-13 keV counts of Cyg X-1 in its soft state with
time step $\Delta t=0.1$s, the dotted curve shows the steady component, i.e. 
the final series of $\{x'\}$
when the fitting process stops, the dashed line is the least-square polynomial of the dotted curve. 
The right panel of Fig. 14
shows the corresponding shot component. Performing the above operation for all the data groups
we obtain a light curve  of the shot component,  $\{x_s\}$, and one of the steady component,
$\{x'\}$, of the observation 
on the time scale $\Delta t$.

With light curves of different components we can calculate characteristic
quantities for each component individually. The hardness ratio distributions 
of different components of Cyg X-1 in different states are shown in Figure 7,
and the variability coherence distributions in  Fig. 9, respectively. 
In Figs. 7 and 9 pluses indicate the average values and their statistical errors 
of the shot component,    circles  the steady component, and
filled circles the original light curve. From these figures 
one can see that although the procedure of distinguishing different components 
we propose is simple, the obtained distributions of two components 
are quite distinct from each other.

\section{DISCUSSION}

Studying variability properties on different time scales directly in the time
domain is an important approach to an understanding of the 
high-energy emission processes in objects. 
With timing analysis in the time domain we can study, in a parallel with the Fourier 
technique, the power spectrum, time lag and coherence of aperiodic variability,
and study other variability properties, e.g. hardness distribution of different
time scales and variability duration, which are difficult to study with
the Fourier technique. 
Although the Fourier technique is a powerful tool  in analyzing periodic and 
quasi-periodic processes,  the Fourier spectrum is not a correct representation
of rms variations in the time domain even for a periodic signal, as shown by Fig.~1.
 As a basis of the Fourier power spectral analysis the Parseval's theorem
\[ \sum_{k=0}^{N-1} |x(t_k)|^2= \frac{1}{N}\sum_{j=-N/2}^{N/2-1} |X(f_j)|^2 \]
simply relates the summed squared modulus of the Fourier amplitudes in the frequency domain 
to the total energy of the process, and says nothing on its energy 
distribution  in the time domain. 
One must be careful in interpreting results of Fourier analyses in the time domain.  
 
Timing directly in the time domain is  useful particularly
in studying short time scale  processes. 
Our results clearly bring out
 the meaningful regularities in the distributions of different quantities studied 
in the short time scale region. For example, characteristic durations significantly
greater than the light curve time steps exist even in short time scale variabilities 
(see Fig. 12). 
The hard X-ray lags (\cite{cui97,now99}) and coherence (\cite{vau97,now99})
resulting  from analyzing $RXTE$ data of Cyg X-1 with the Fourier transform technique     
provide important diagnostics of the accretion region and emission mechanism. 
In the high frequency region of $> 30 - 50$ Hz, however,  the errors in the  time lags 
and coherence obtained from the Fourier analysis are too large to use in quantitatively
model investigates.
Our results of hardness, coherence and time lags from the analysis technique
in the time domain, shown in Fig.~7, Fig.~9  and Fig.~11, have good statistics in the short 
time scale region as well, which can be taken as an observational basis for
studying emission and propagation processes on millisecond time scales. 

In comparison with the Fourier technique timing in the time domain has the freedom of 
choosing proper statistics.
For example, we use $H$ (16) and $r_d$ (18)  to describe hardness and
coherence  respectively  as opposed to using $h$ (15) and $r_x$ (17),
which have undesirable structures for 
fake light curves of constant mean intensity and pure Poisson statistics.
The timing technique in the time domain has no rigorous requirement for
continuity of  data. The final value of a characteristic quantity on  a given time scale 
in the time domain results from values of different data groups being statistically averaged.
Continuity in observation is required only for a single group which should be longer than the  
time scale under study, it doesn't matter how long and how many data gaps between 
two used data groups.  
We can, in principal, use the technique in the time domain to derive statistically meaningful
results in the study of rapid variabilities of weak sources, e.g. AGNs, by synthesizing 
data of different observations for a certain object.   

The distributions of different characteristic 
quantities vs. time scale of Cyg X-1 in different states reveal a variety 
of  important properties of the X-ray emission process. 
It seems that the whole region of  time scales studied can be divided into three 
regions roughly with the two division points $\Delta t\sim$ 0.01~s and 0.1~s.
  The distinguishing features of time scale dependence of variability properties,
e.g. signal power density  (Fig.~3), 
 hard X-ray lag (Fig.~11), variation width (Figs.~12, 13),  
harness ratio and coherence of shot component (Fig.~7 and Fig.~9) 
are often different in the time scale regions of $\Delta t < 0.01$~s, between 0.01 s and
0.1~s and $\Delta t >$0.1~s.
At all states of Cyg X-1 in the short time scale region of $\Delta t<0.01$ s 
as the time scale decrease,  the power density of X-ray variability 
decreases and coherence becomes weaker. There are some variability properties
 showing state dependent: 
e.g.  in the short time scale region around $\Delta t \sim 0.01$ s 
the variability duration of high-energy photons 
is shorter than that of low-energy photons in the transition states,
but wider in the hard state, as is shown in Fig. 13.
These properties should be considered in the construction of emission models. 

It is obvious that a simple Comptonization model, i.e. hard X-rays come from 
multiple inverse Compton scattering of thermal photons in electron gas with high 
temperature, can not explain many of the temporal characteristics revealed by
timing analysis in the time domain. 
The variability duration in high-energy band being shorter in the  
transition states of Cyg X-1 is  
in contradiction to the expectation from the Comptonization process.
A natural explanation is that some seed photons of short pulses have energies already
in the hard X-ray region and the property of high-energy pulses having shorter duration is an
intrinsic character of seed pulses. It seems that the effect of 
spectrum, temporal property and production mechanism of such seed photons on the observed hard X-rays 
 can not be ignored.  The duration of short variability in high-energy band of Cyg X-1 
in the hard state is, as distinguished from the other states and just as expected by
 the Comptonization process, wider than
that in low-energy band.  According to the advection dominated accretion flow (ADAF)
model, a large region of ADAF with high temperature around the center black hole
exists only in the hard state (see, e.g. \cite{esi97}). Therefore the effect of
Comptonization process should be most significant in the hard state.
If this is true, then the difference of variation durations between  the hard state and the other
states reflects the geometry of the accretion region.
In the long time scale region of $\Delta t>0.05$ s, high-energy time delay increases
with time scale up to above 10 ms, which is greater than the radial light crossing,
sound crossing or free-fall time scales in ADAF (this result is consistent with 
time lags in low Fourier frequency region from the Fourier spectrum analysis, see 
\cite{now99}).  Besides propagation in ADAF, other processes, i.e. heating in accretion
disc or ADAF,  should be considered in interpreting hard X-ray delay on long time scale.
In summary, production of hard X-rays from black hole binaries may be related to 
geometric structure and physical properties of accretion region, as well as to
multiple processes with different time scales.   The timing technique in the time
domain is an useful tool to expose their specific properties that are hard 
to be revealed by the Fourier technique.

The analysis technique in the time domain is far from being completed, it needs 
further developing in various aspects. We try to distinguish
different components of emission with a simple procedure in the time domain.
The fact that  the shot and steady  components resulted from our procedure have
quite different distributions of hardness and coherence (see Fig. 7 and Fig. 9), 
indicates that such a distinguishing is meaningful: they more or less 
correspond to different processes with different characters.   
In comparison with the values from total light curve, the hardness ratios of the shot component 
of Cyg X-1 on the short time scale of $\Delta t < 0.01 - 0.1$ s are always lower in the hard state 
and higher in other states, and the variability coherence of shot component
is always weaker.  The hardness and coherence of the steady component in each 
state of Cyg X-1 are near to the values from the total light curve.
These results should be conducive to 
searching and investigating different physical processes involved in X-ray emission.
    
%\acknowledgments 
The author thanks the referee for helpful comments and Yu Wenfei,  
Chen Li and Qu Jinglu for discussions.  
This research was supported by the Special Funds for Major State Basic
Research Projects and by the National Natural
Science Foundation of China and  made use 
of data obtained through the High Energy Astrophysics Science Archive Research 
Center Online Service,
provided by the NASA/Goddard Space Flight Center.

\begin{figure}
\epsscale{1.0}
\vspace{5cm}
\plotfiddle{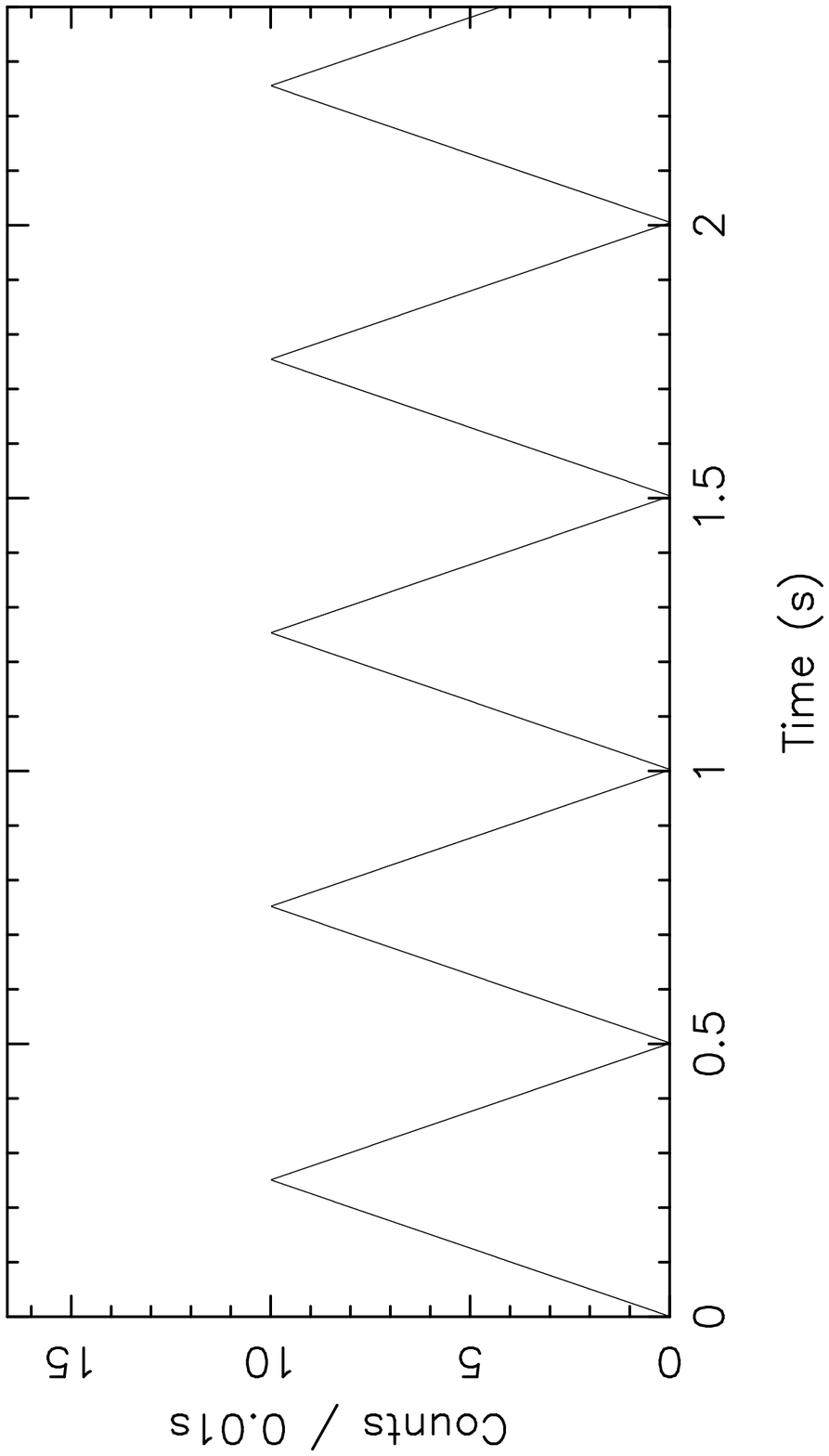}{20pt}{-90}{40}{20}{-200}{220}
\plotfiddle{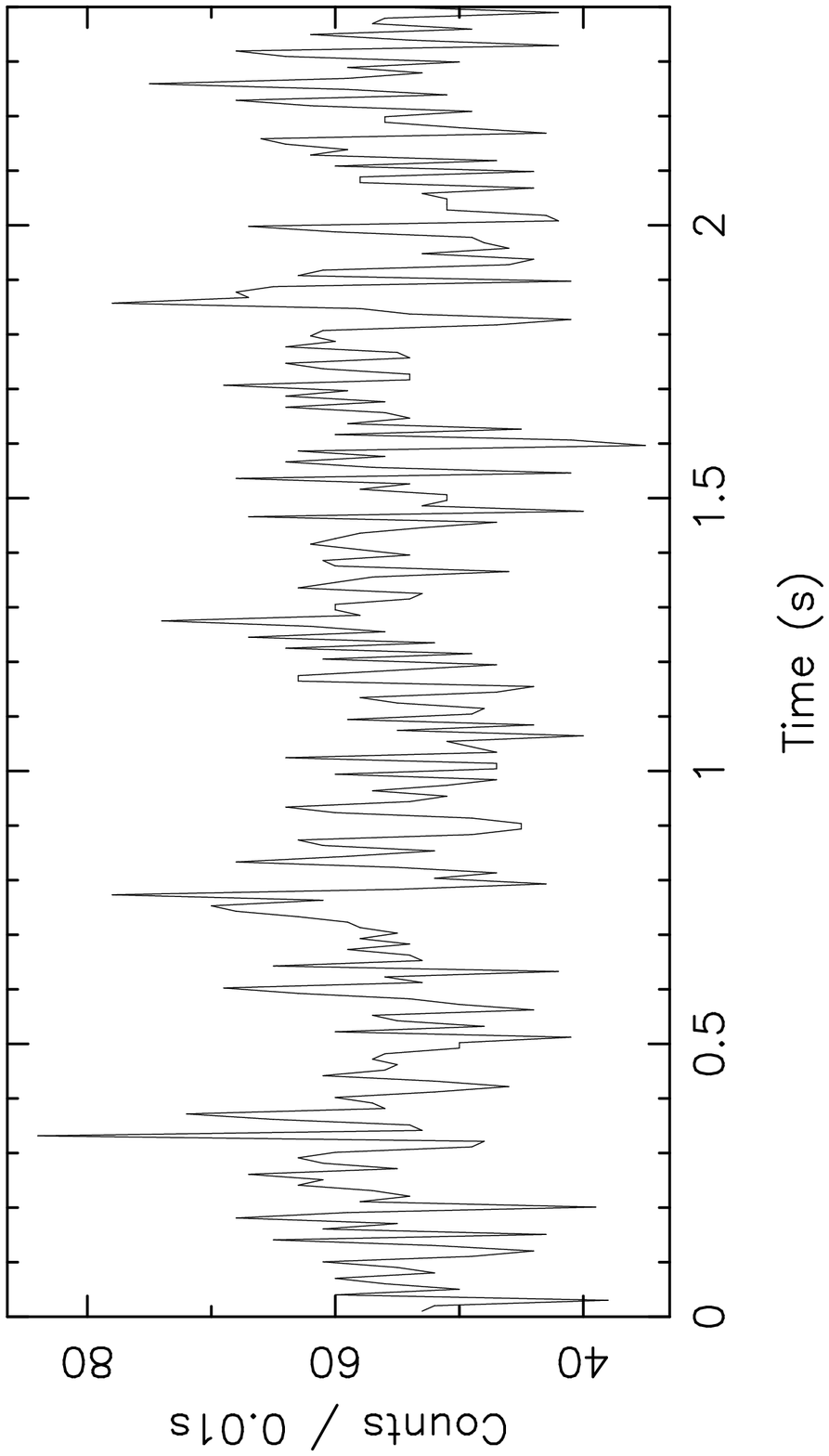}{20pt}{-90}{40}{25}{-200}{180}
\vspace{1.8cm}
\plotfiddle{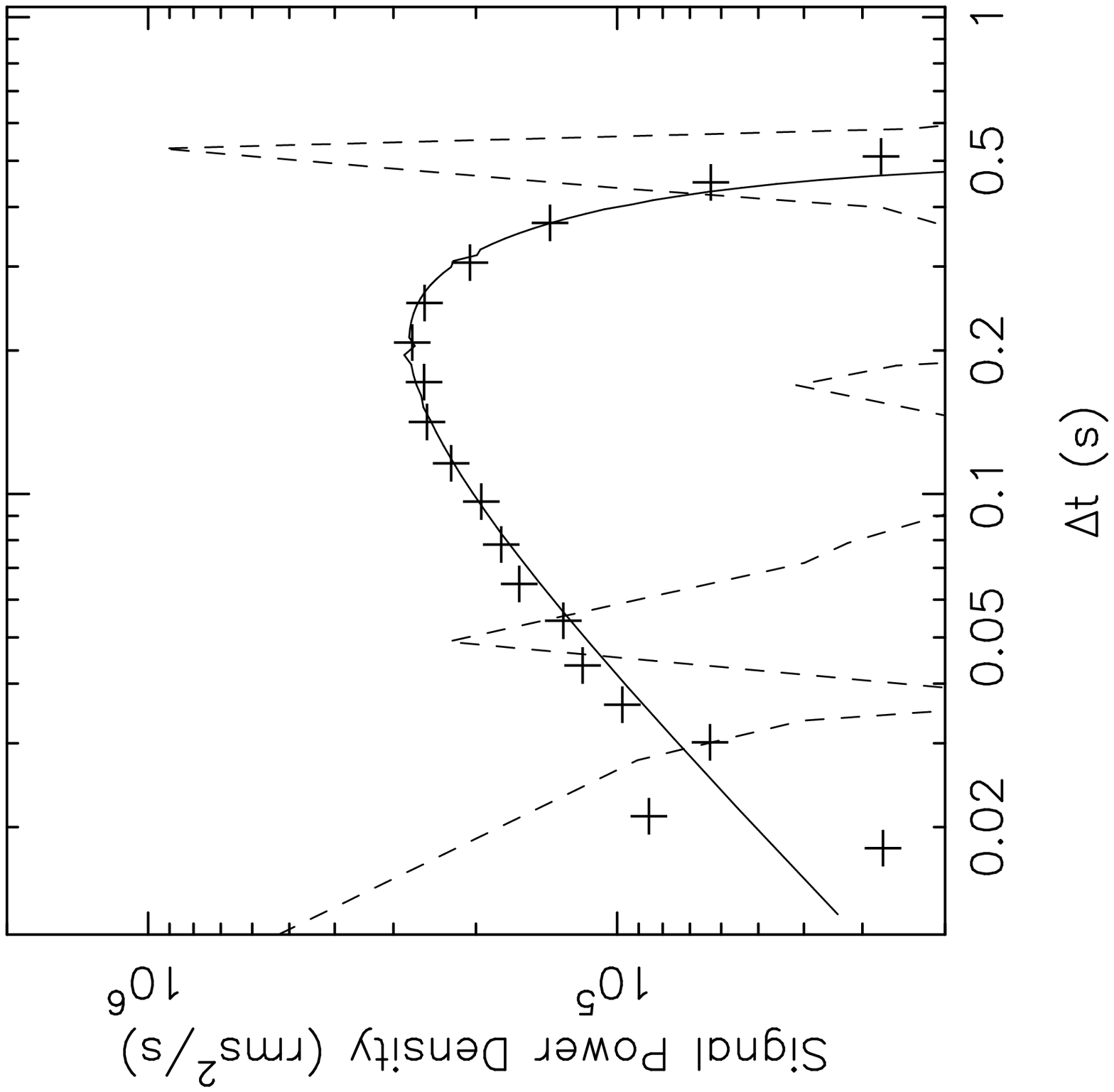}{20pt}{-90}{50}{48}{-190}{130}
\vspace{5cm}
\caption{Distribution of power density vs. time scale of a periodic triangle signal. 
{\sl Top panel}: the signal. {\sl Middle panel}: simulated data = signal + Poisson noises.
{\sl Bottom panel}: signal power densities . {\it Solid line} -- theoretical distribution
of variation power densities expected by the signal. 
{\it Dashed line} -- excess Fourier spectrum after subtracting the noise spectrum.
{\it Plus} -- excess power densities
calculated by the timing technique in the time domain. 
\label{fig1}}
\end{figure}

\begin{figure}
\epsscale{1.0}
\vspace{5cm}
\plotfiddle{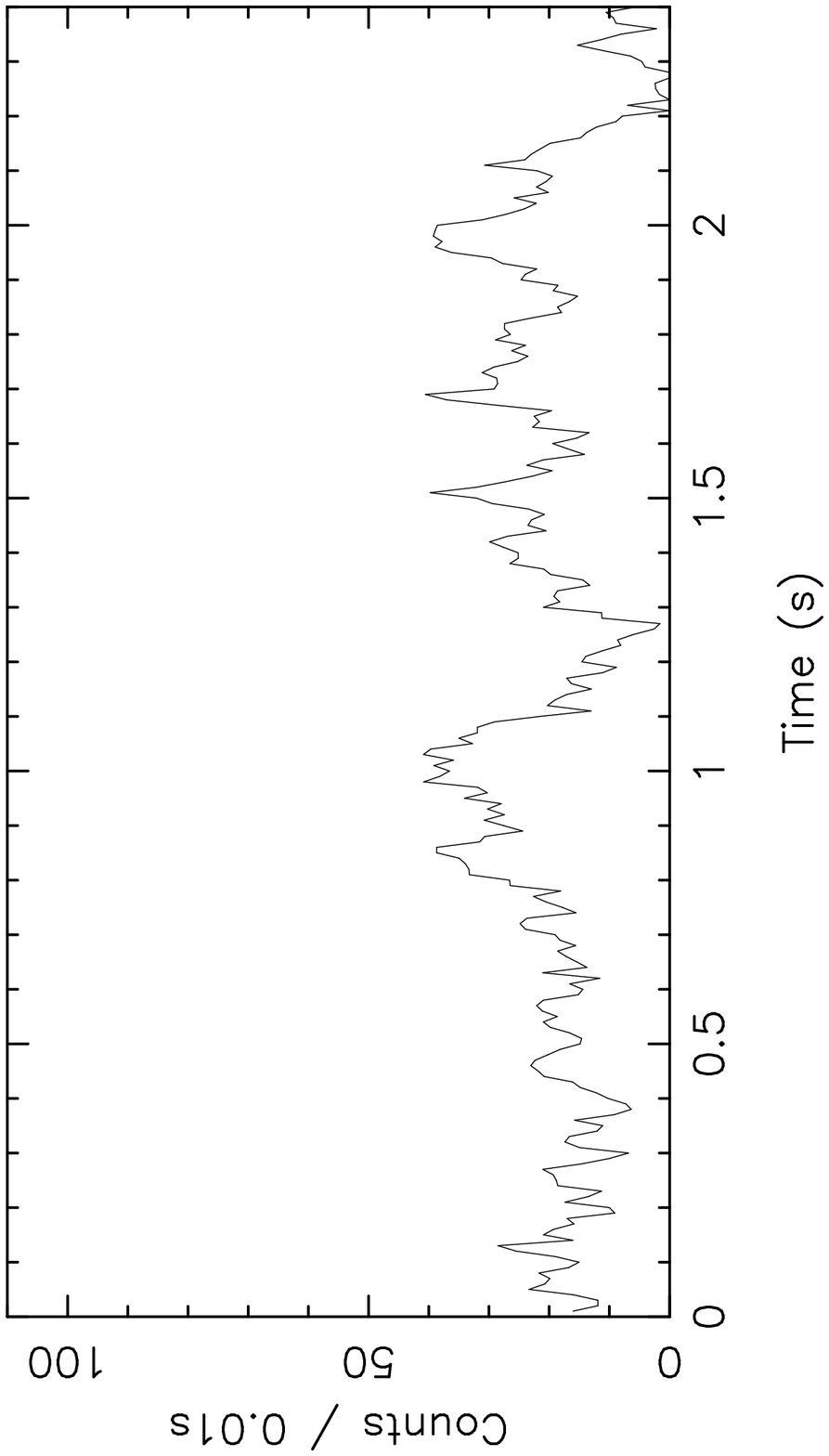}{20pt}{-90}{40}{20}{-200}{220}
\plotfiddle{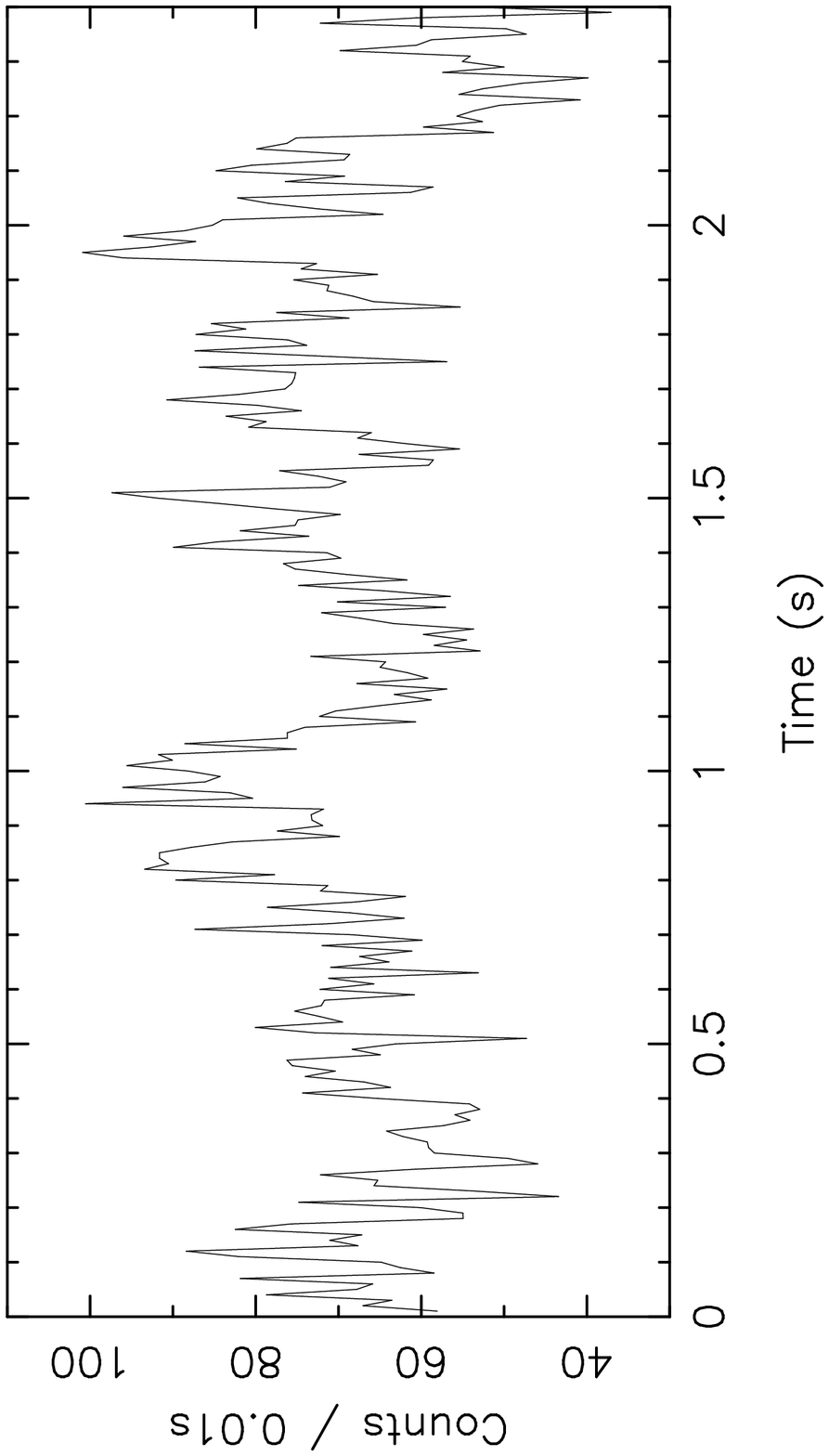}{20pt}{-90}{40}{25}{-200}{180}
\vspace{1.8cm}
\plotfiddle{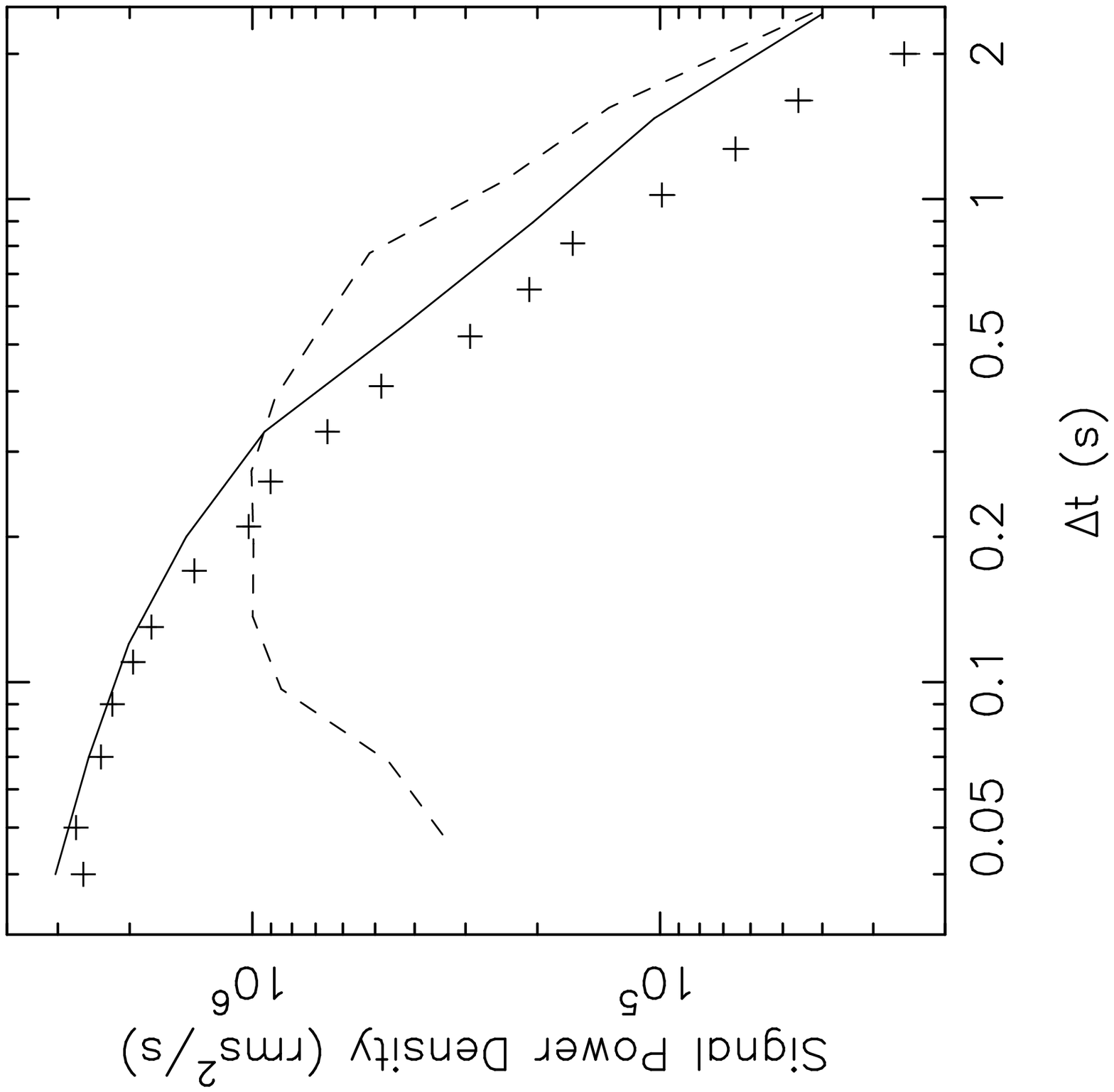}{20pt}{-90}{50}{48}{-190}{130}
\vspace{5cm}
\caption{Distribution of power density vs. time scale of a signal of AR process. 
{\sl Top panel}: the signal. {\sl Middle panel}: simulated data = signal + Poisson noises.
{\sl Bottom panel}: signal power densities . {\it Solid line} -- theoretical distribution
of variation power densities expected by the signal. 
{\it Dashed line} -- excess Fourier spectrum.
{\it Plus} -- excess power densities
calculated by the timing technique in the time domain. 
\label{fig2}}
\end{figure}

\begin{figure}

\vspace{4cm}
\epsscale{1.0}
\plotfiddle{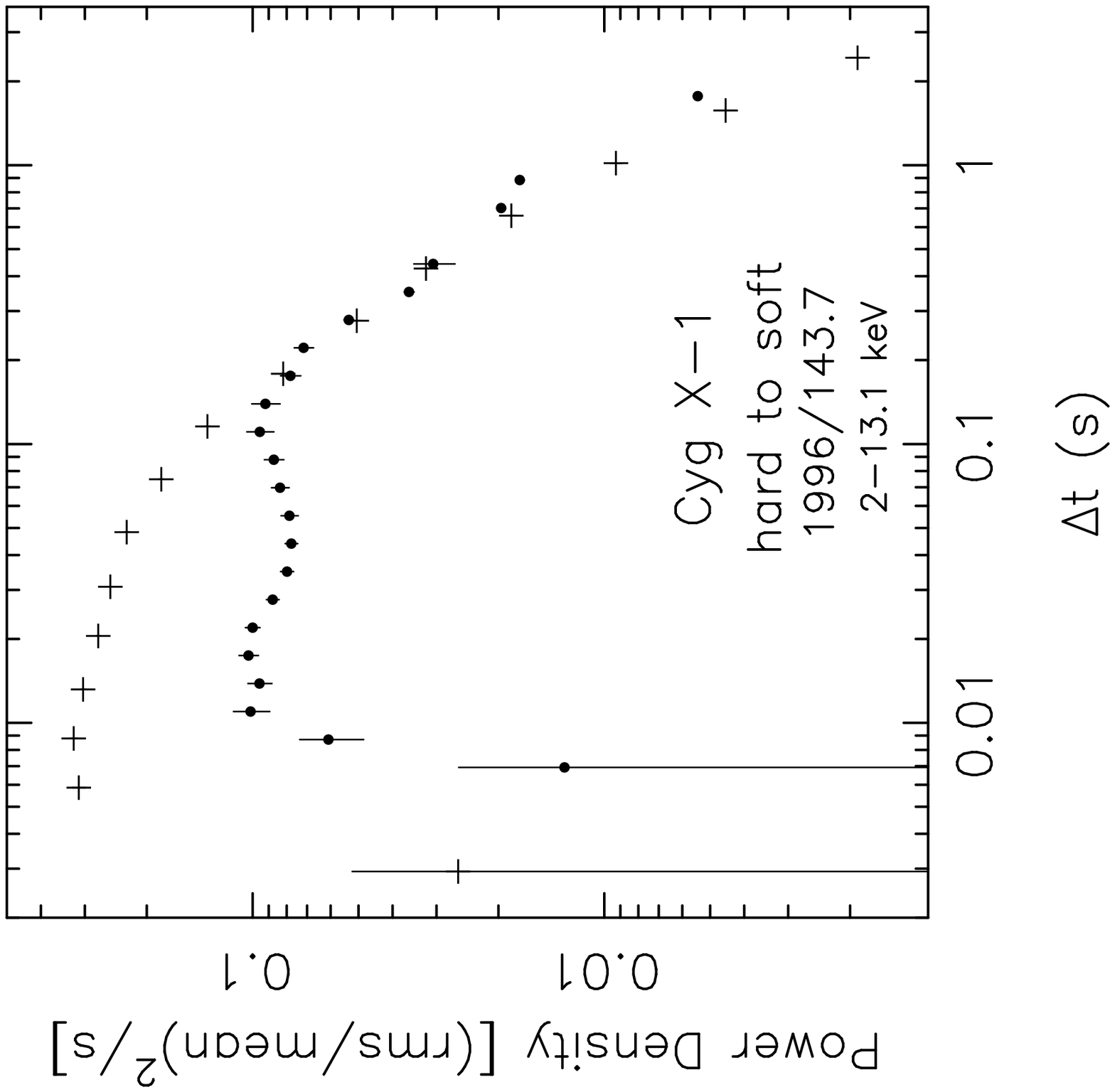}{20pt}{-90}{40}{25}{-260}{178}
\plotfiddle{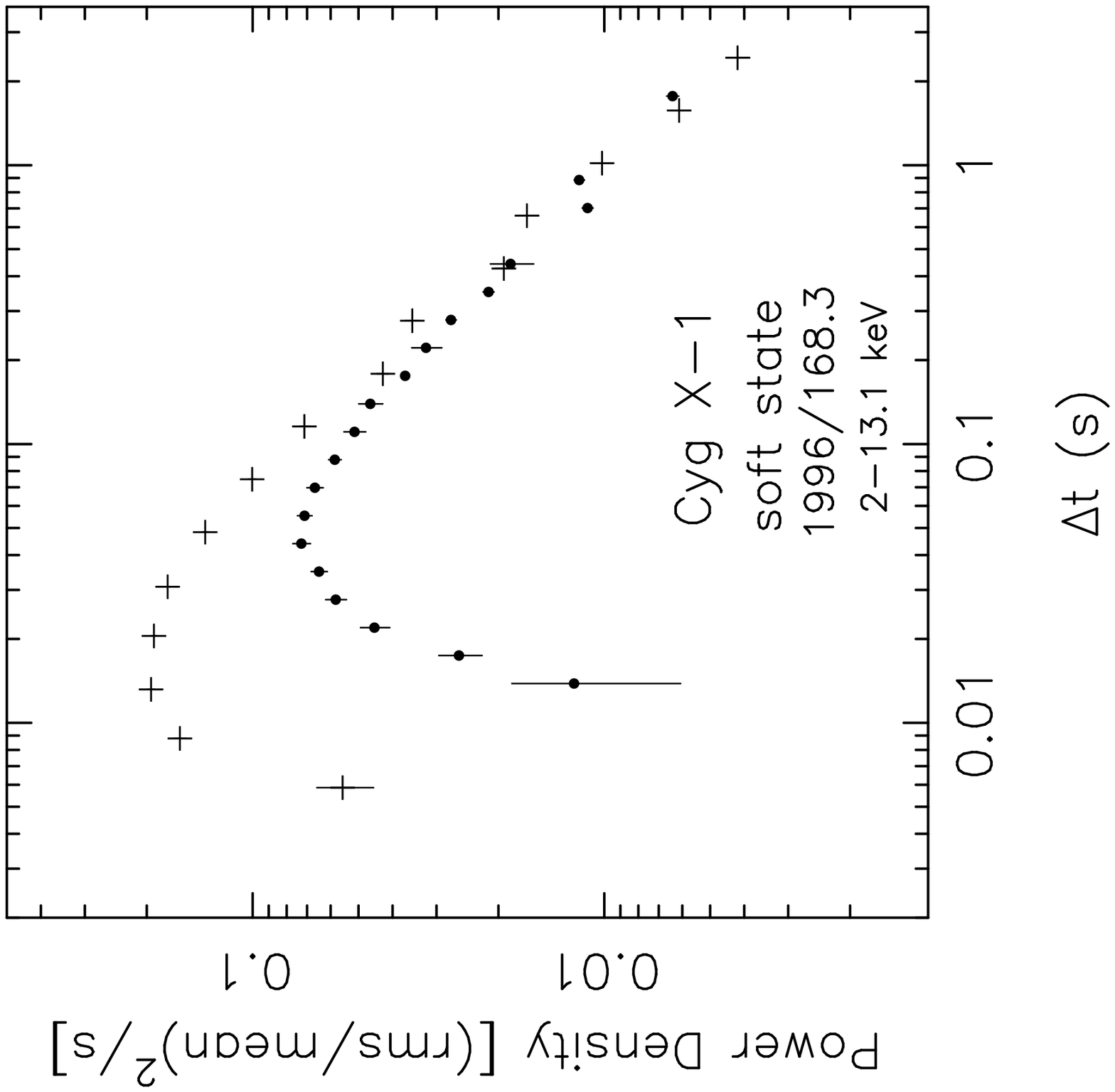}{20pt}{-90}{40}{25}{-50}{212}

\vspace{2.5cm}
\plotfiddle{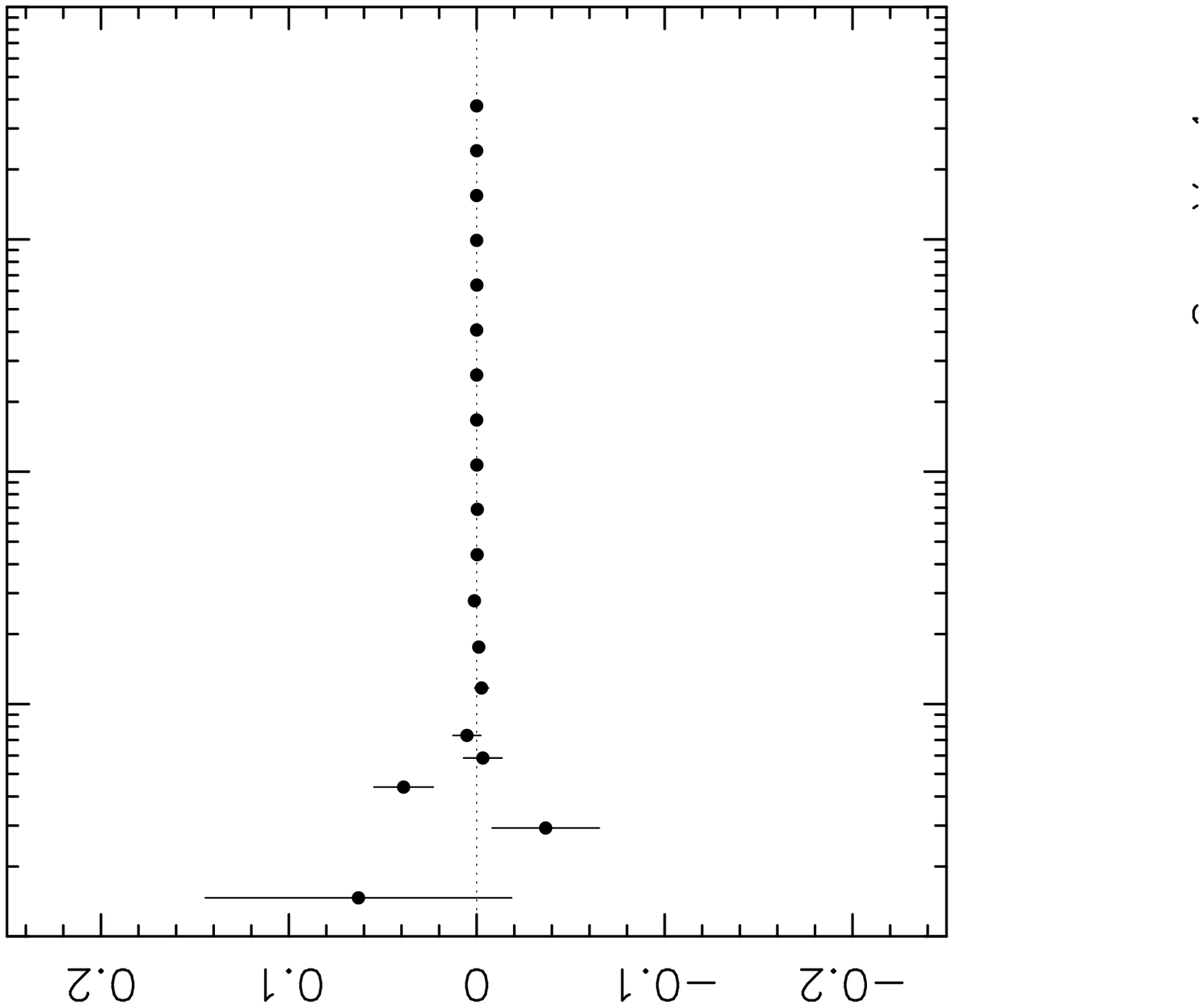}{20pt}{-90}{40}{25}{-260}{178}
\plotfiddle{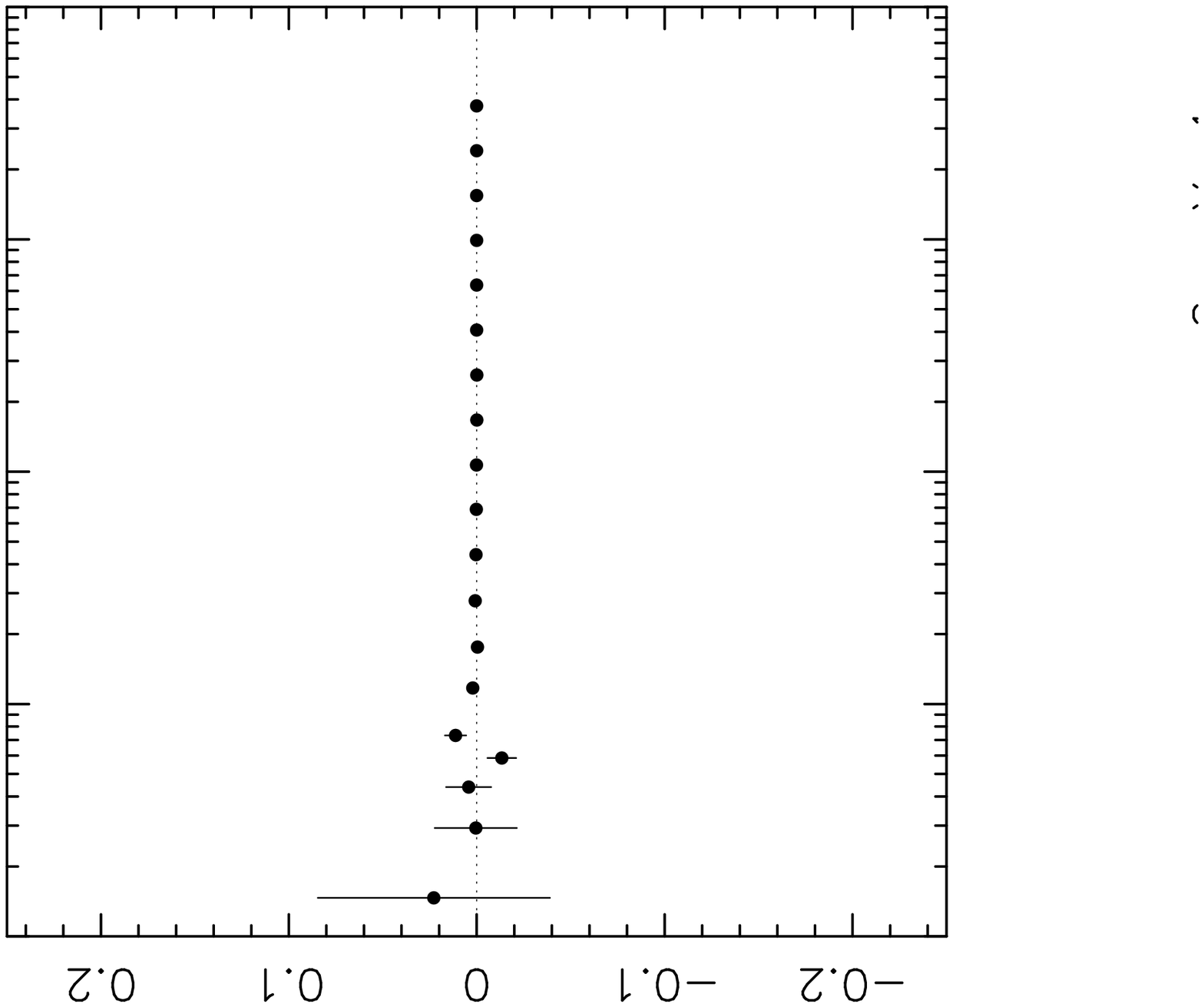}{20pt}{-90}{40}{25}{-50}{212}

%\vspace{-1.5cm}
\plotfiddle{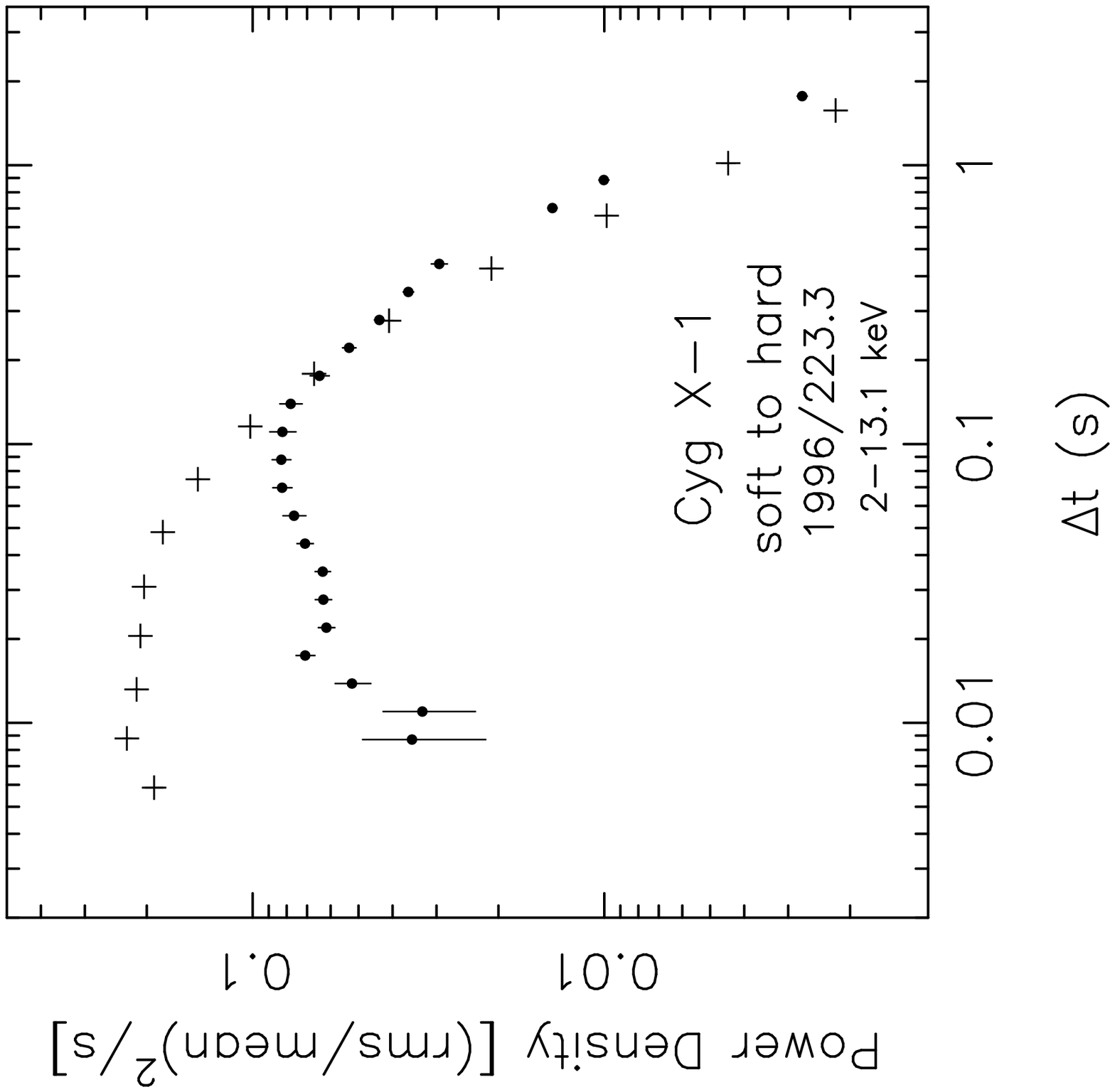}{20pt}{-90}{40}{25}{-260}{118}
\plotfiddle{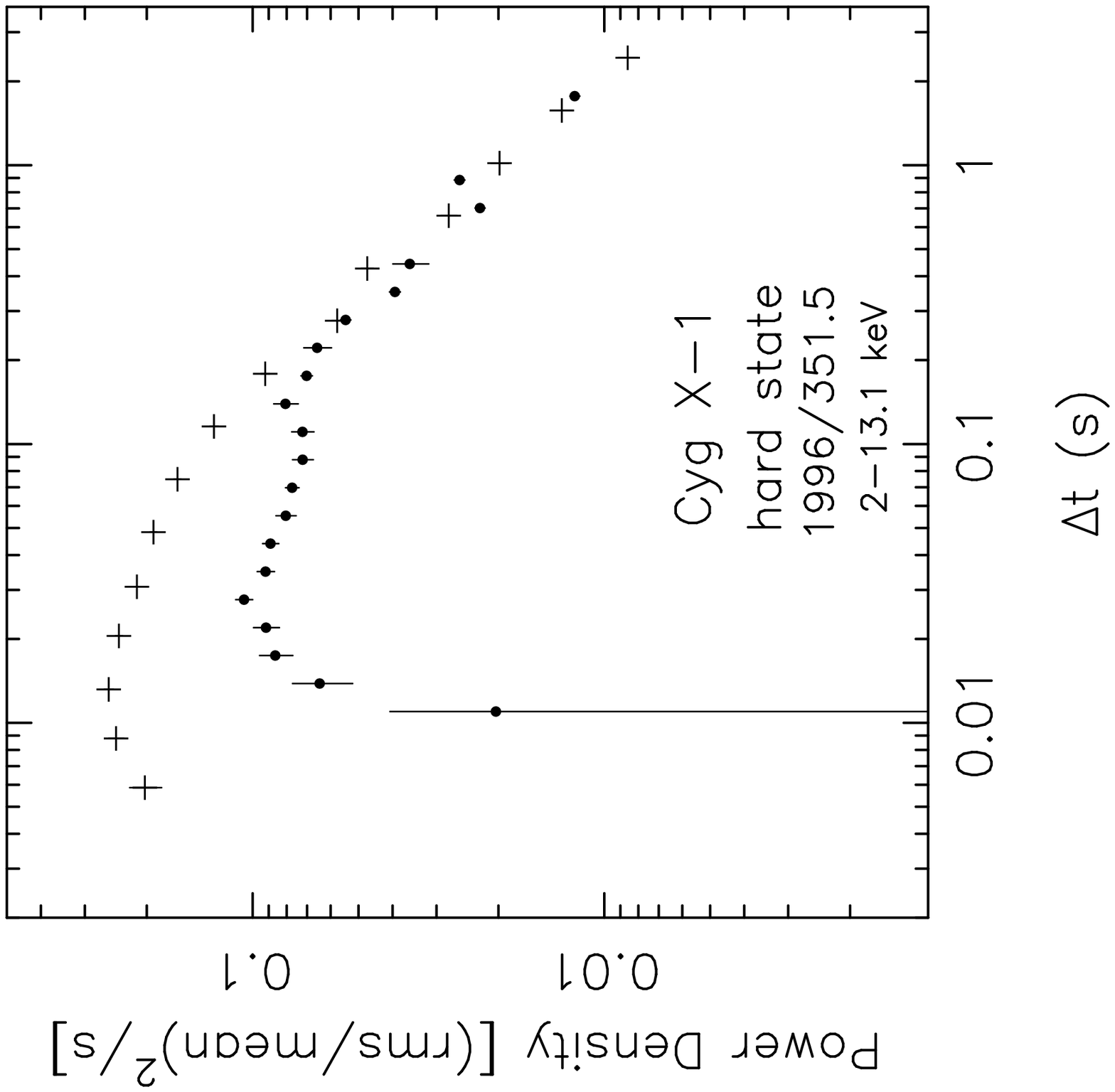}{20pt}{-90}{40}{25}{-48}{152}

\vspace{4.5cm}
\plotfiddle{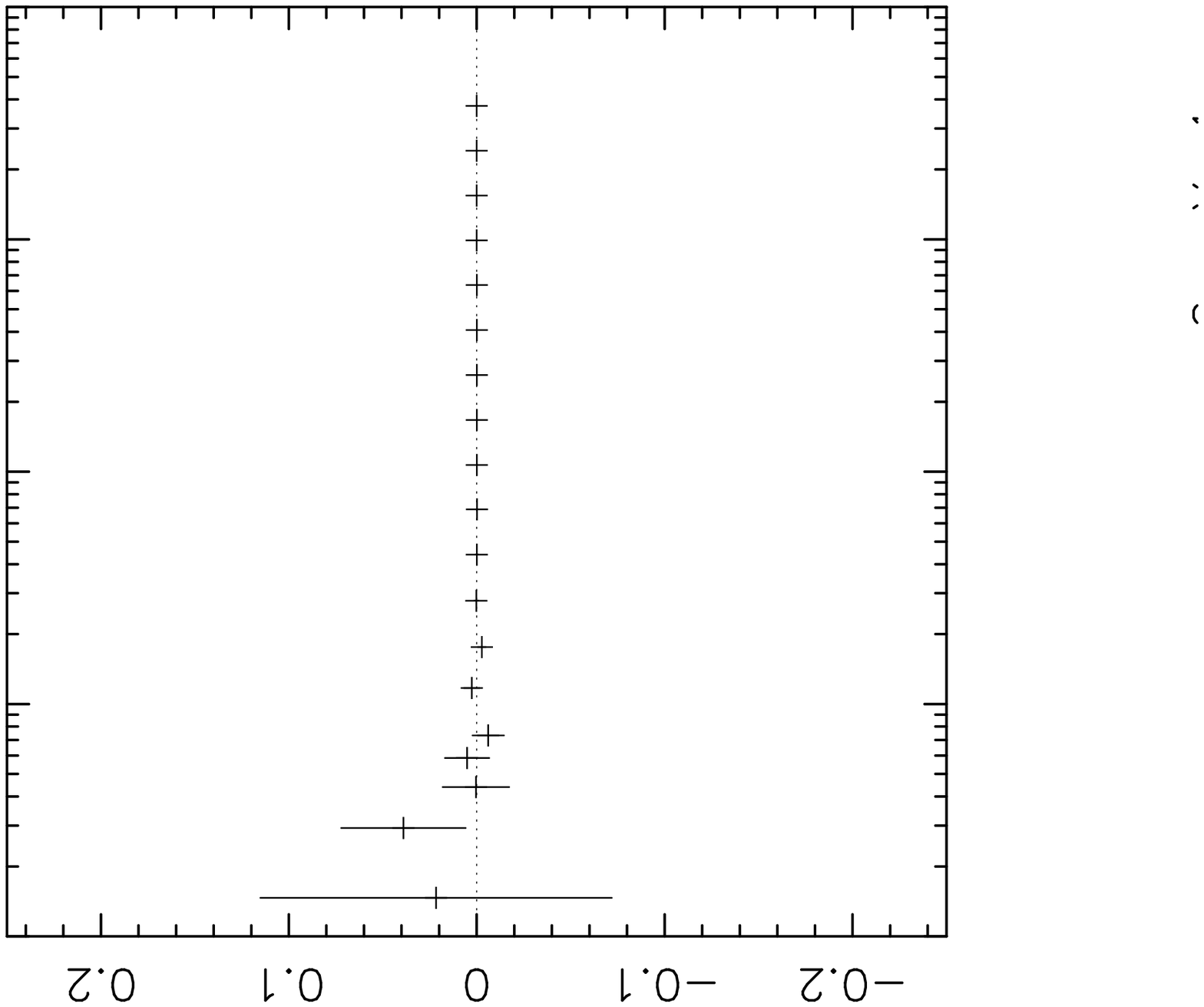}{20pt}{-90}{40}{15}{-260}{178}
\plotfiddle{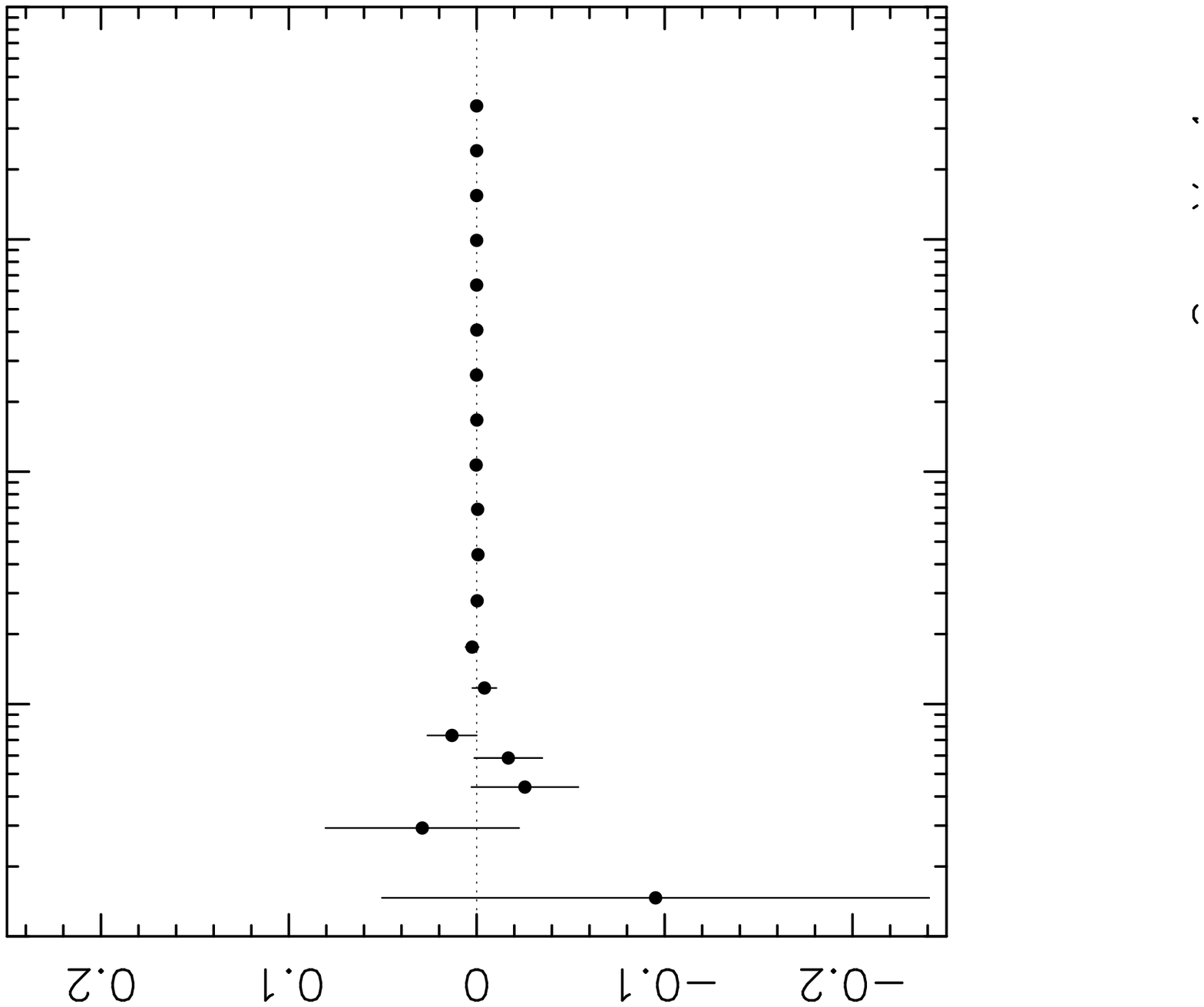}{20pt}{-90}{40}{15}{-50}{212}

\vspace{-4cm}
\caption{Fractional signal power density vs. time scale of Cyg X-1 in different state
calculated directly in the time domain ({\it plus}) and through the Fourier transformation
 ({\it dot}).
The narrow panel under each plot of Cyg X-1 shows 
the fractional signal power densities calculated in the time domain 
for a fake light curve of Poisson noise with mean as 
the average counting rate of the observed Cyg X-1 light curve. 
\label{fig3}}
\end{figure}

\begin{figure}
\vspace{2.5cm}
\epsscale{1.0}
\plotfiddle{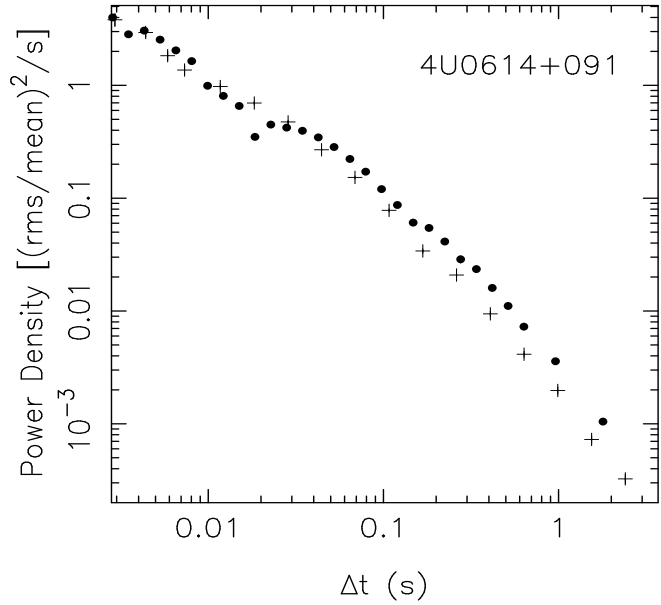}{20pt}{-90}{50}{45}{-180}{178}
\vspace{2.5cm}
\caption{Fractional signal power density vs. time scale of the neutron star binary 
4U0614+091 $3-20$ keV X-rays
analyzed in the time domain ({\it plus}) and in the frequency domain
 ({\it dot}). The analyzed data was observed by PCA/$RXTE$ on April 30, 1998. 
\label{fig4}}

\end{figure}

\begin{figure}
\vspace{2.5cm}
\epsscale{1.0}
\plotfiddle{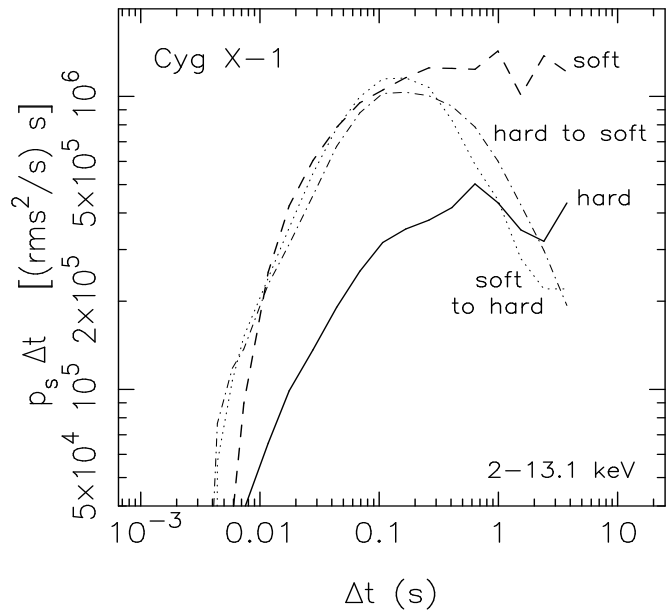}{20pt}{-90}{50}{45}{-180}{178}
\vspace{2.5cm}
\caption{Distributions of signal power density $\bar{p}_s$ multiplied by time scale $\Delta t$
of different states of Cyg X-1.  \label{fig5}}
\end{figure}

\begin{figure}
\vspace{5.5cm}
\epsscale{1.0}
\plotfiddle{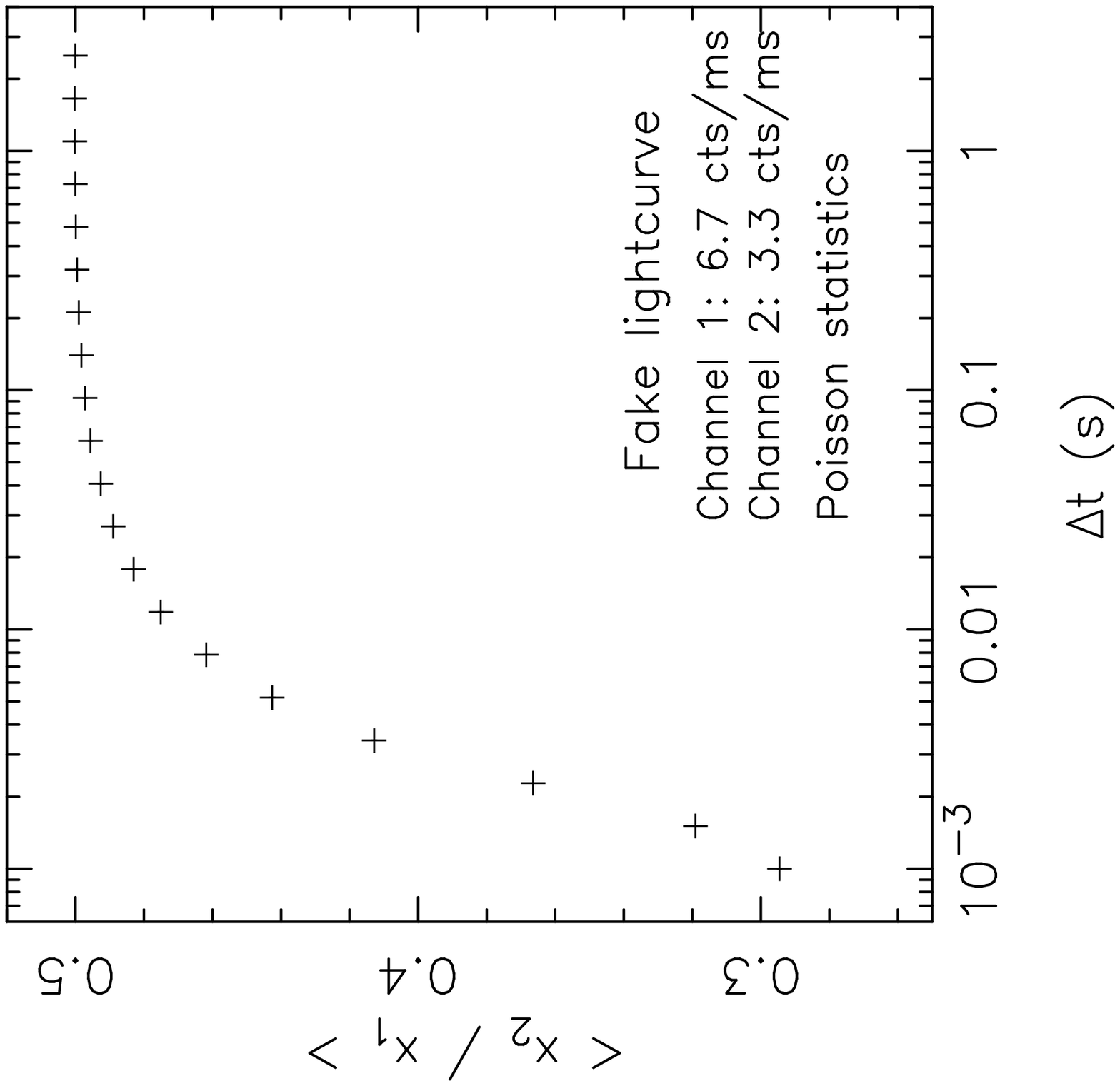}{20pt}{-90}{40}{25}{-260}{178}
\plotfiddle{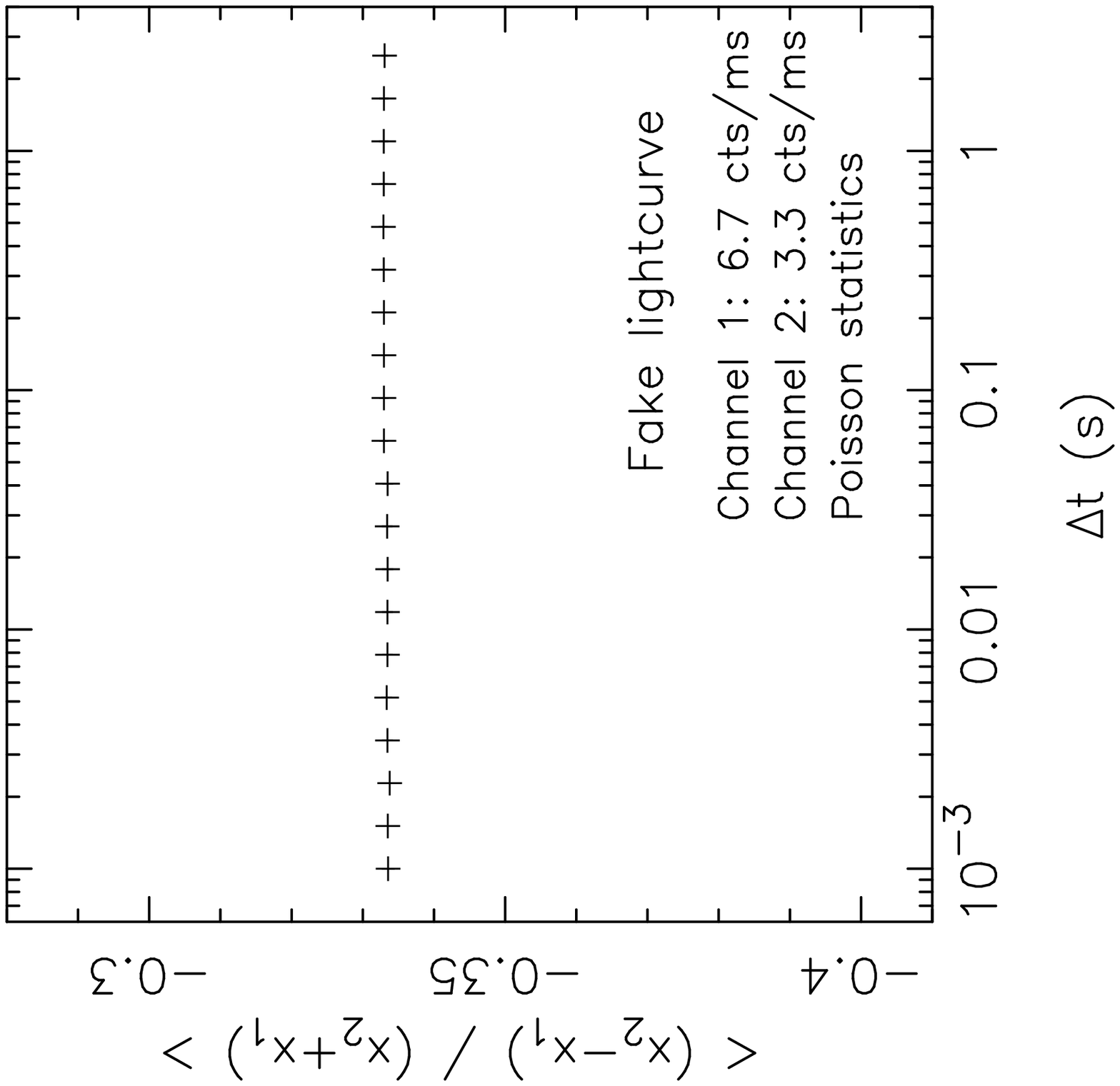}{20pt}{-90}{40}{25}{-50}{212}
\vspace{-26mm}
\caption{Hardness vs. time scale from fake light curves
\label{fig6}}
\end{figure}

 \begin{figure}
\vspace{4.5cm}
\epsscale{1.0}
\plotfiddle{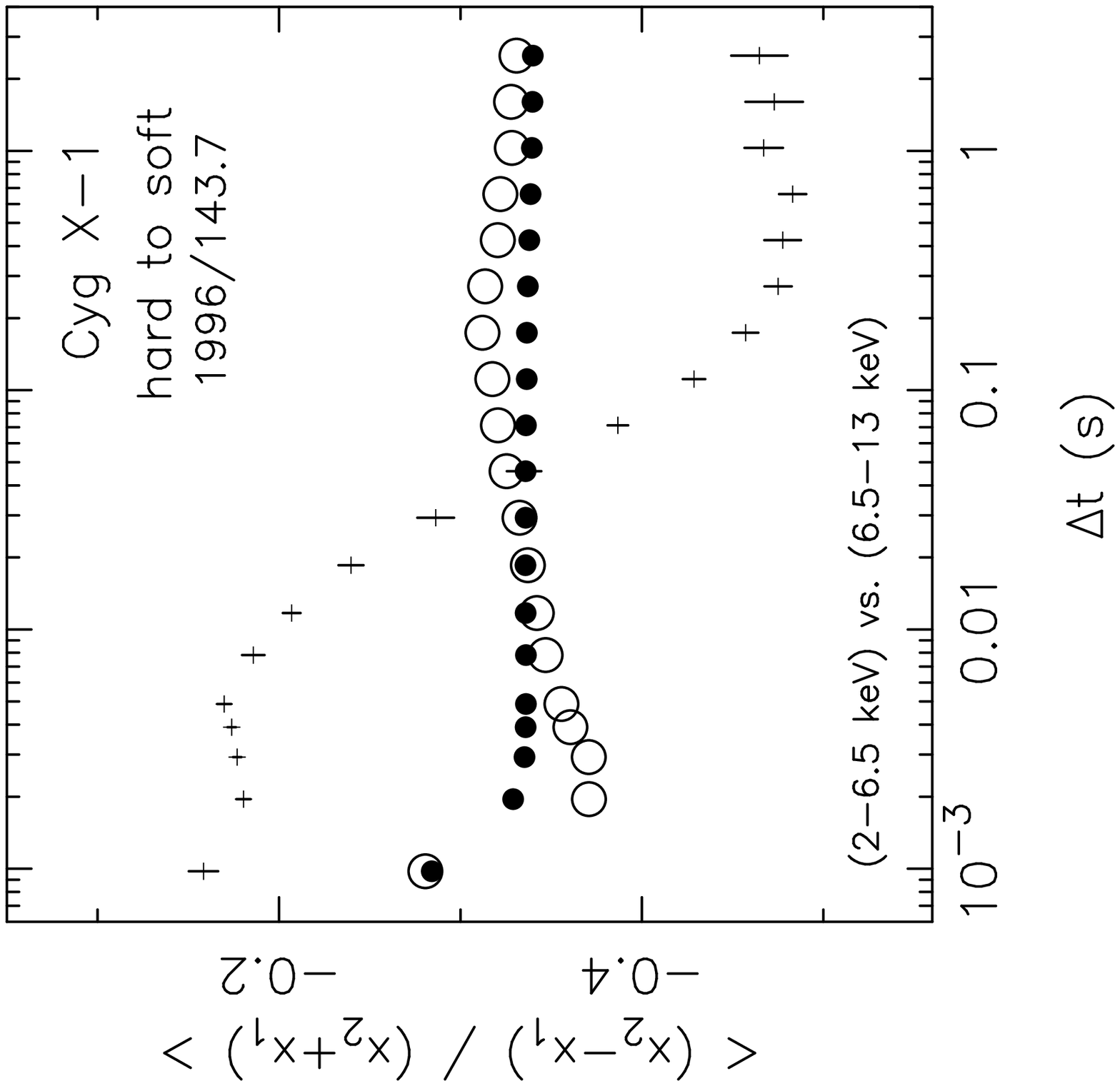}{20pt}{-90}{40}{25}{-260}{178}
\plotfiddle{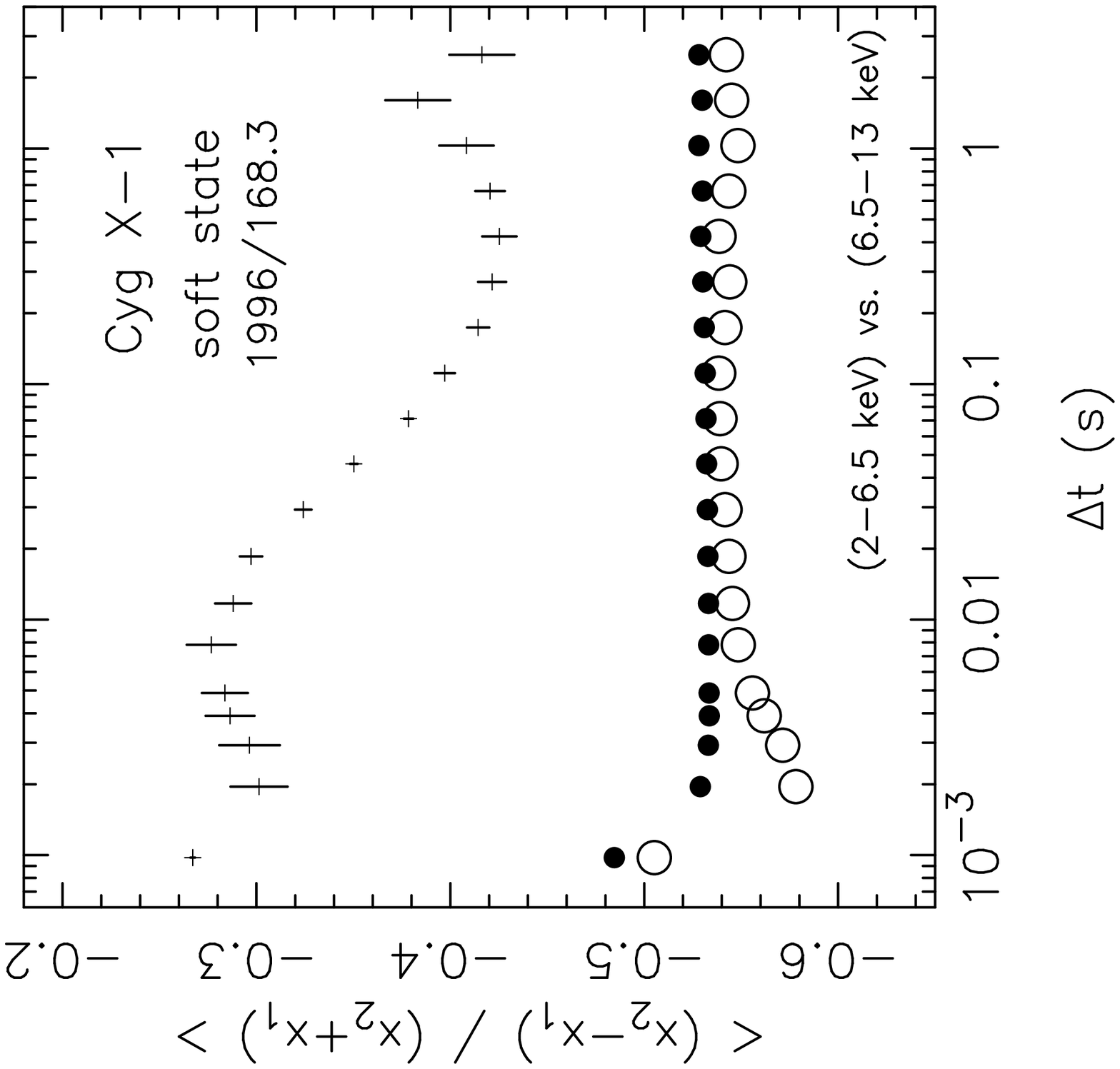}{20pt}{-90}{40}{25}{-50}{212}
\plotfiddle{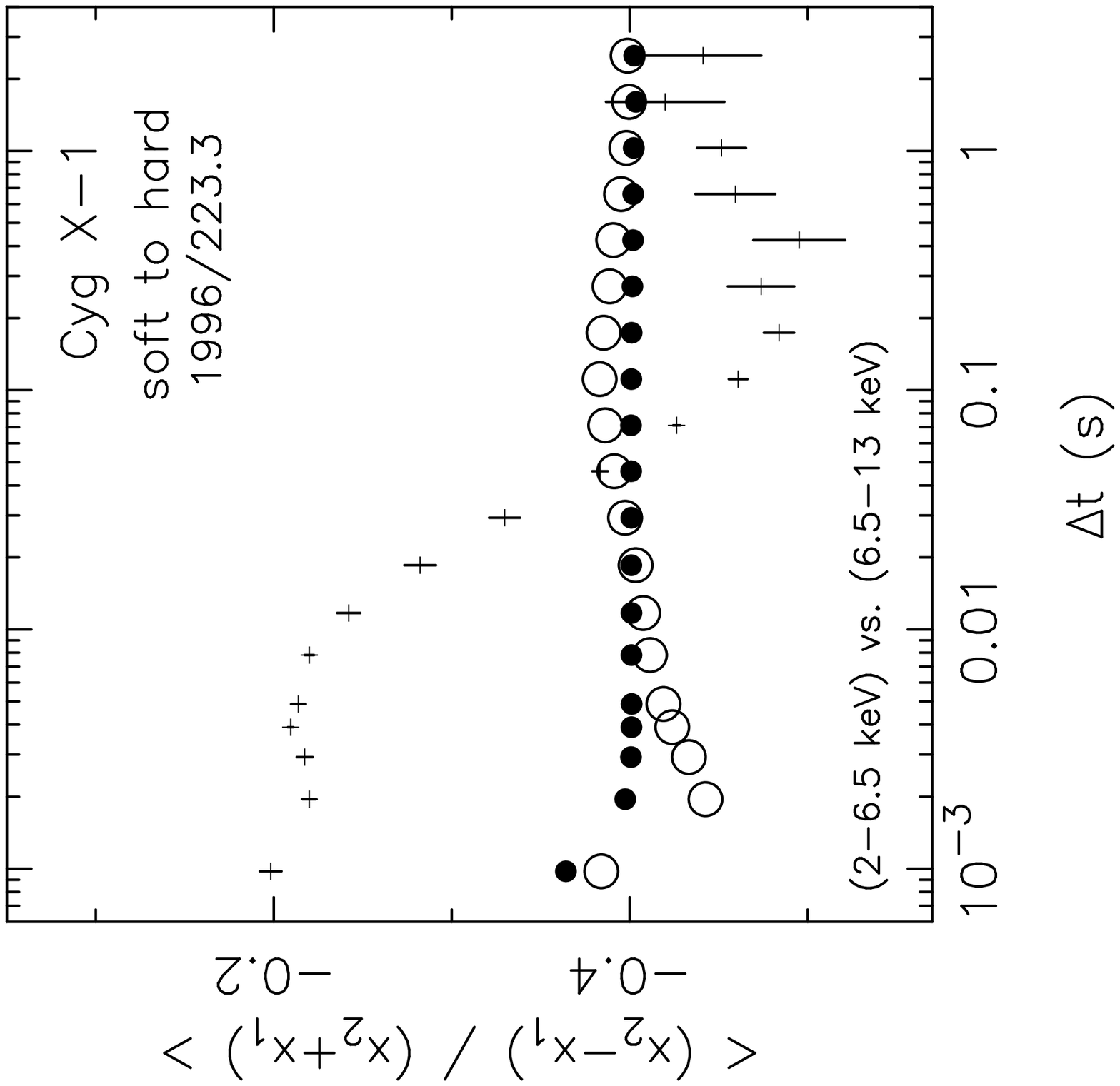}{20pt}{-90}{40}{25}{-260}{118}
\plotfiddle{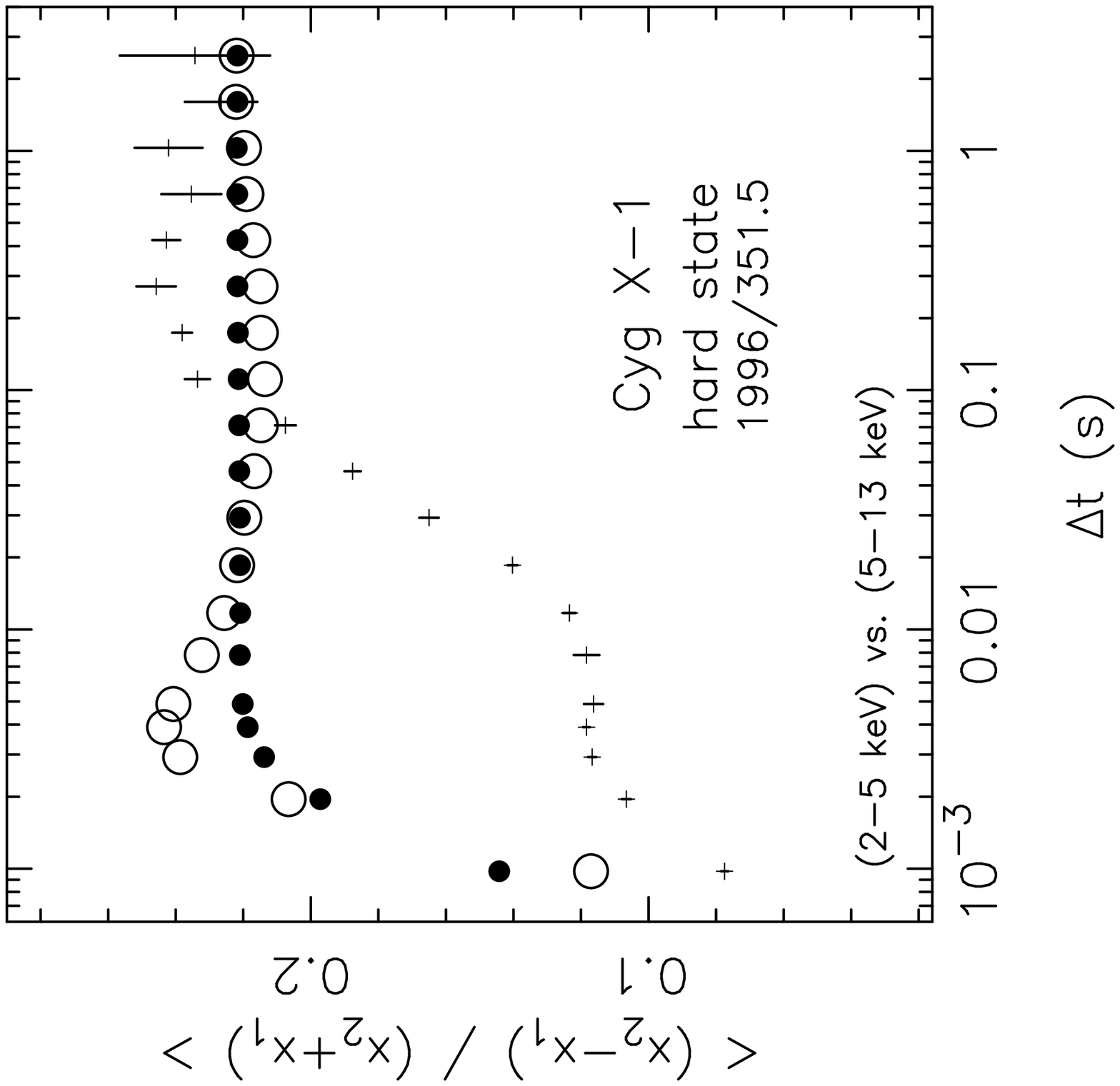}{20pt}{-90}{40}{25}{-50}{152}
\vspace{-4mm}
\caption{Hardness ratio of different component vs. time scale of Cyg X-1.
  {\it Filled circle}: total light curve. {\sl Plus}: shot component.  {\it Circle}: steady component. 
  \label{fig7}}
\end{figure}

 \begin{figure}
\vspace{5cm}
\epsscale{1.0}
\plotfiddle{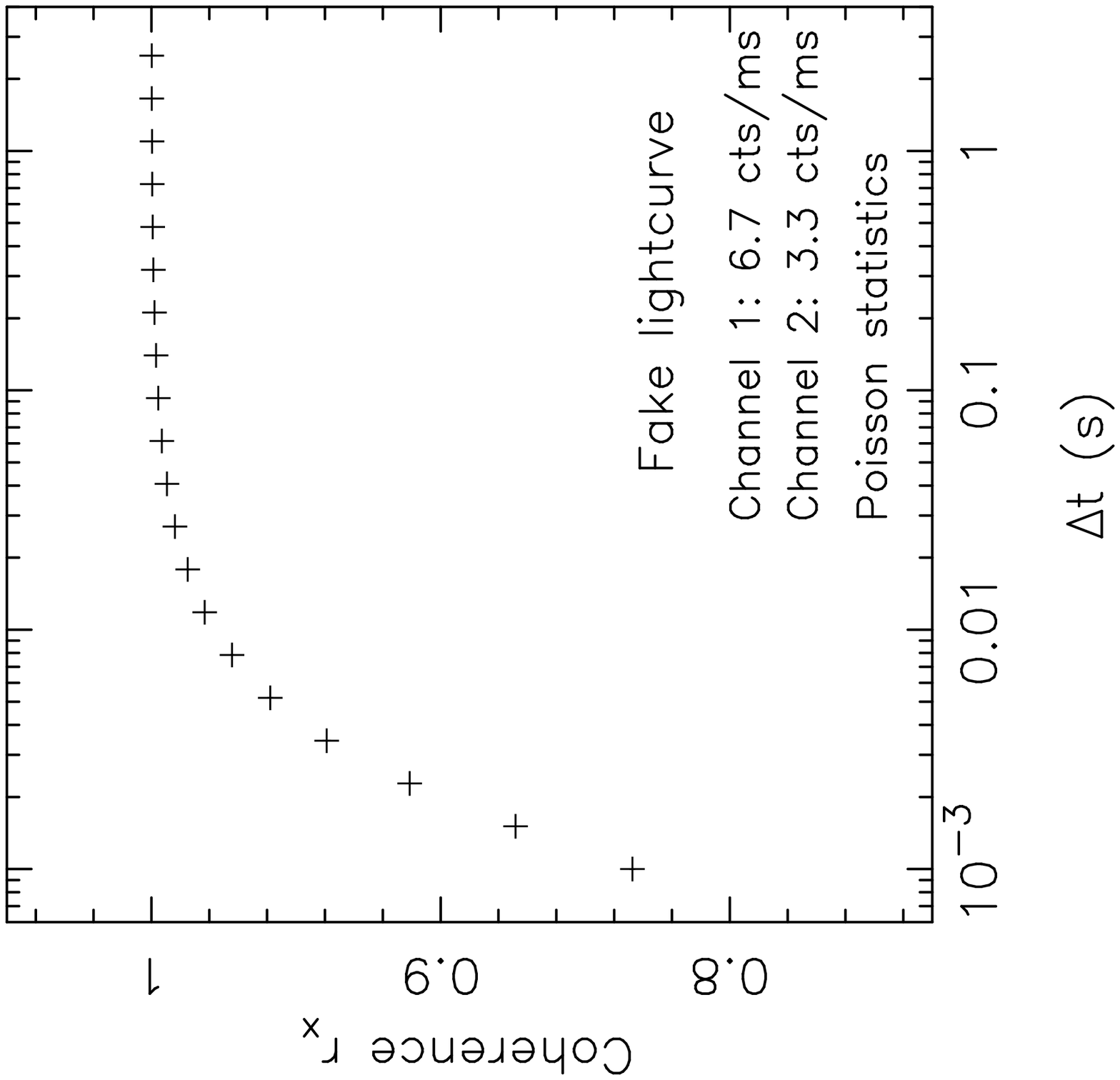}{20pt}{-90}{40}{25}{-260}{178}
\plotfiddle{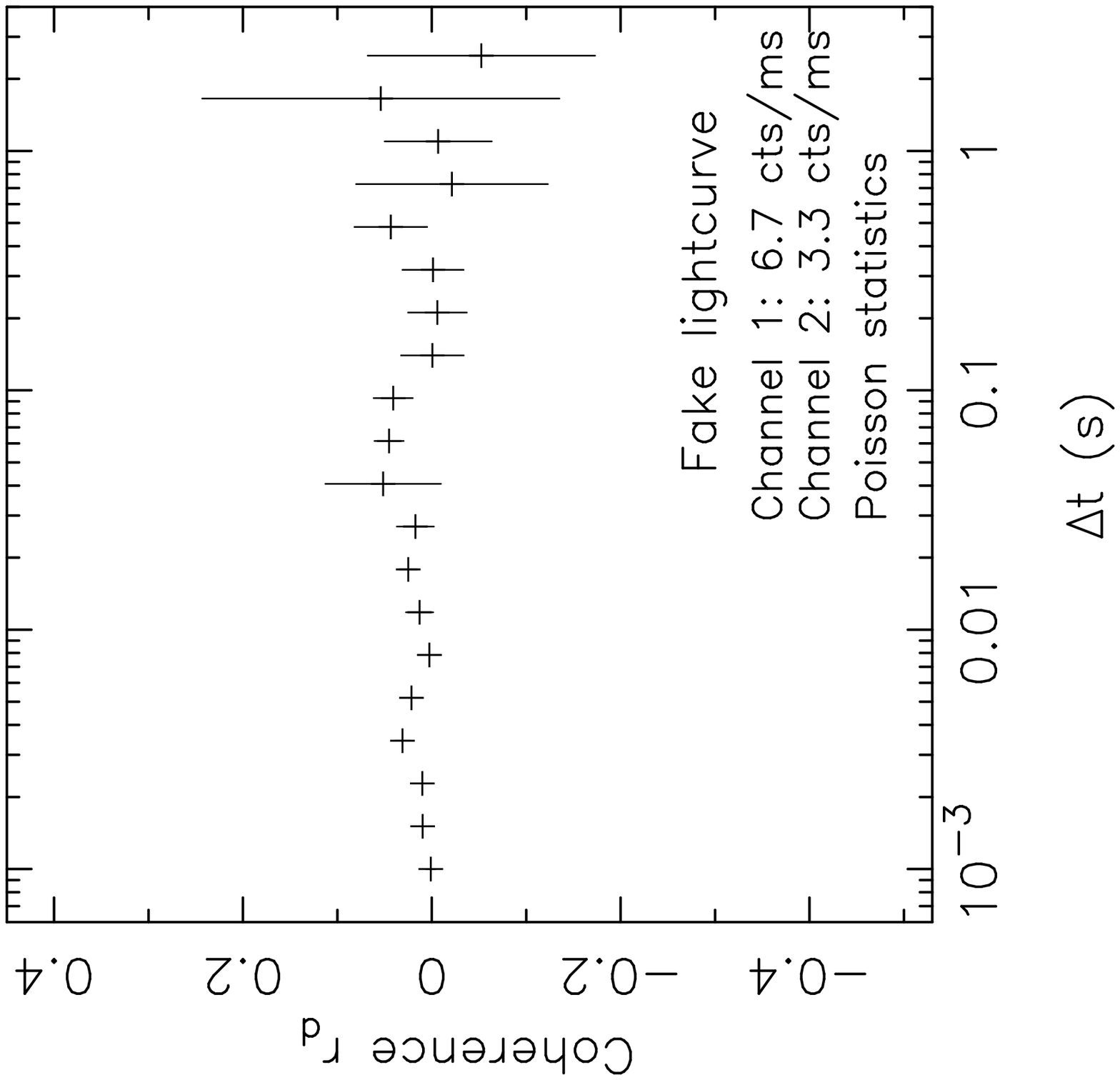}{20pt}{-90}{40}{25}{-50}{212}
\vspace{-21mm}
\caption{Coherence vs. time scale of fake light curves  \label{fig8}}
\end{figure}

 \begin{figure}
\vspace{4.8cm}
\epsscale{1.0}
\plotfiddle{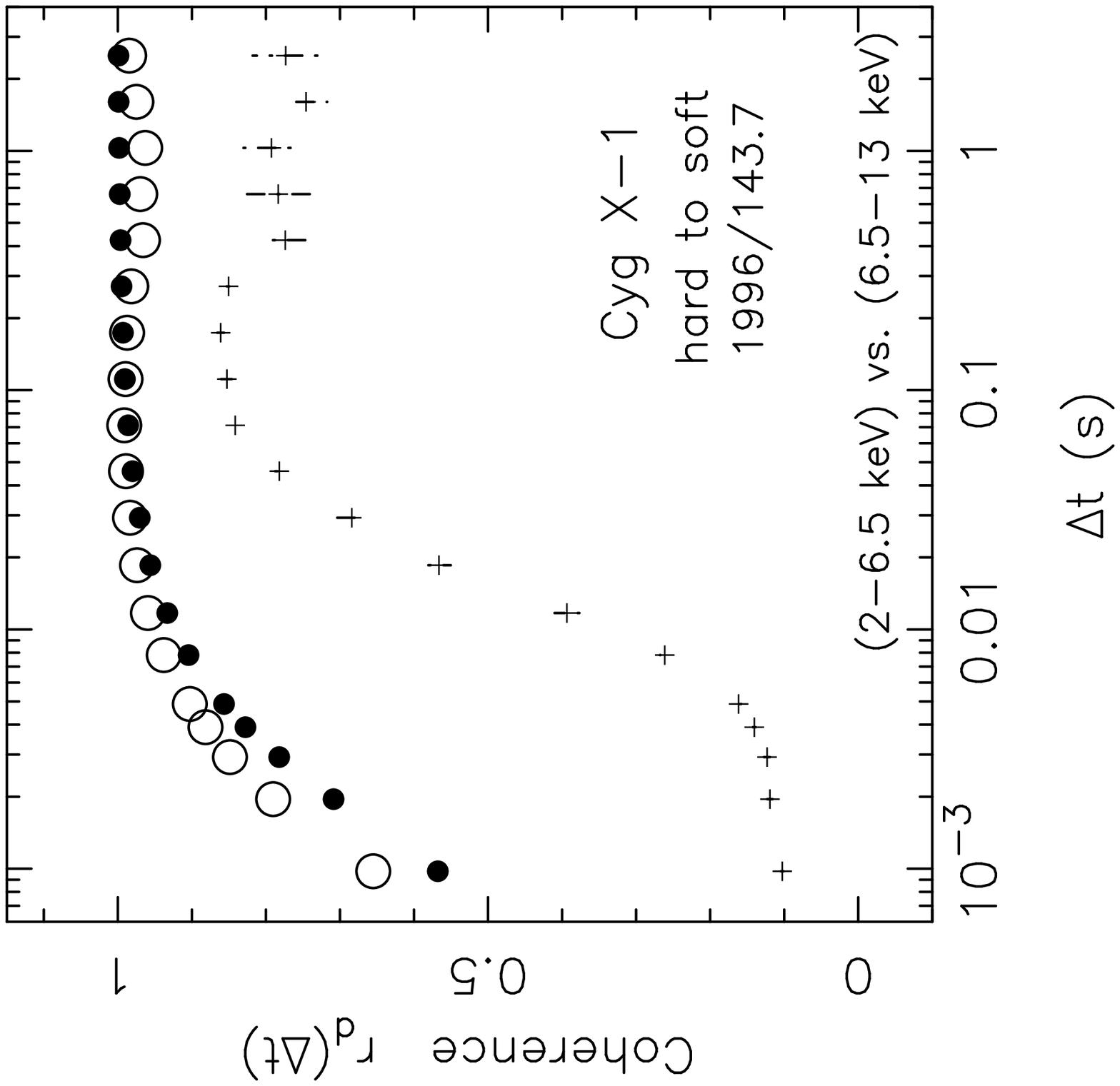}{20pt}{-90}{40}{25}{-260}{178}
\plotfiddle{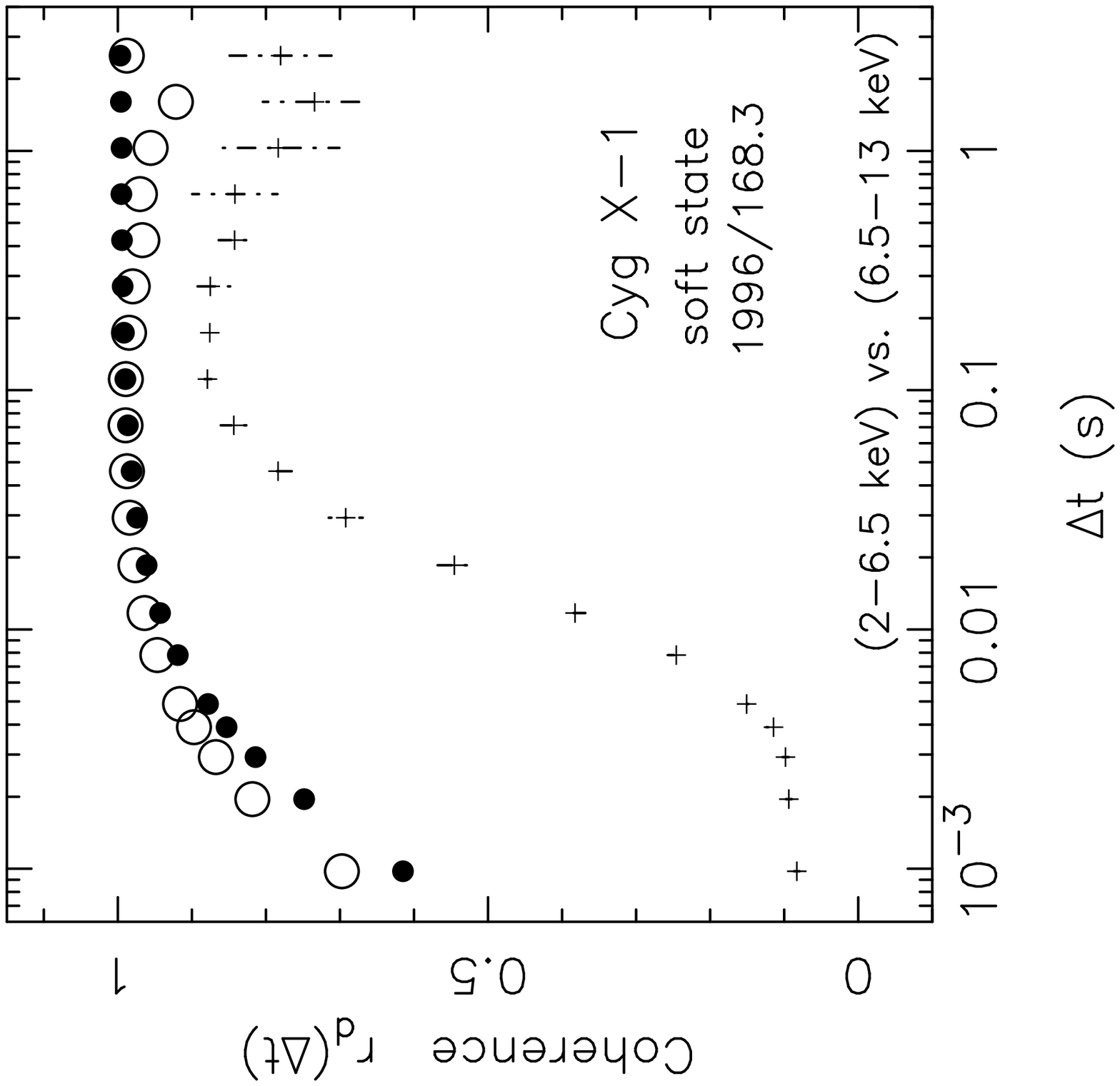}{20pt}{-90}{40}{25}{-50}{212}
\plotfiddle{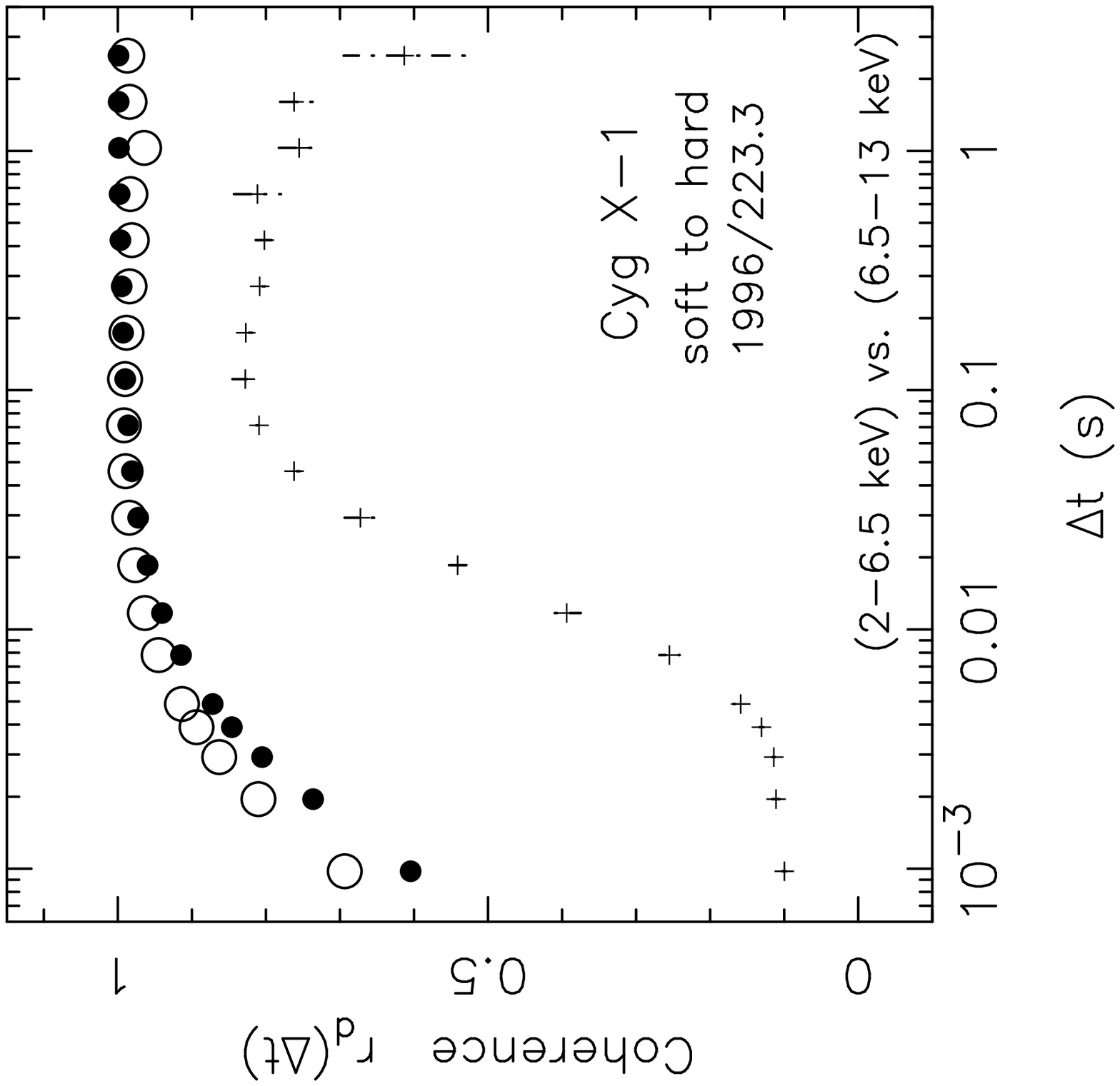}{20pt}{-90}{40}{25}{-260}{119}
\plotfiddle{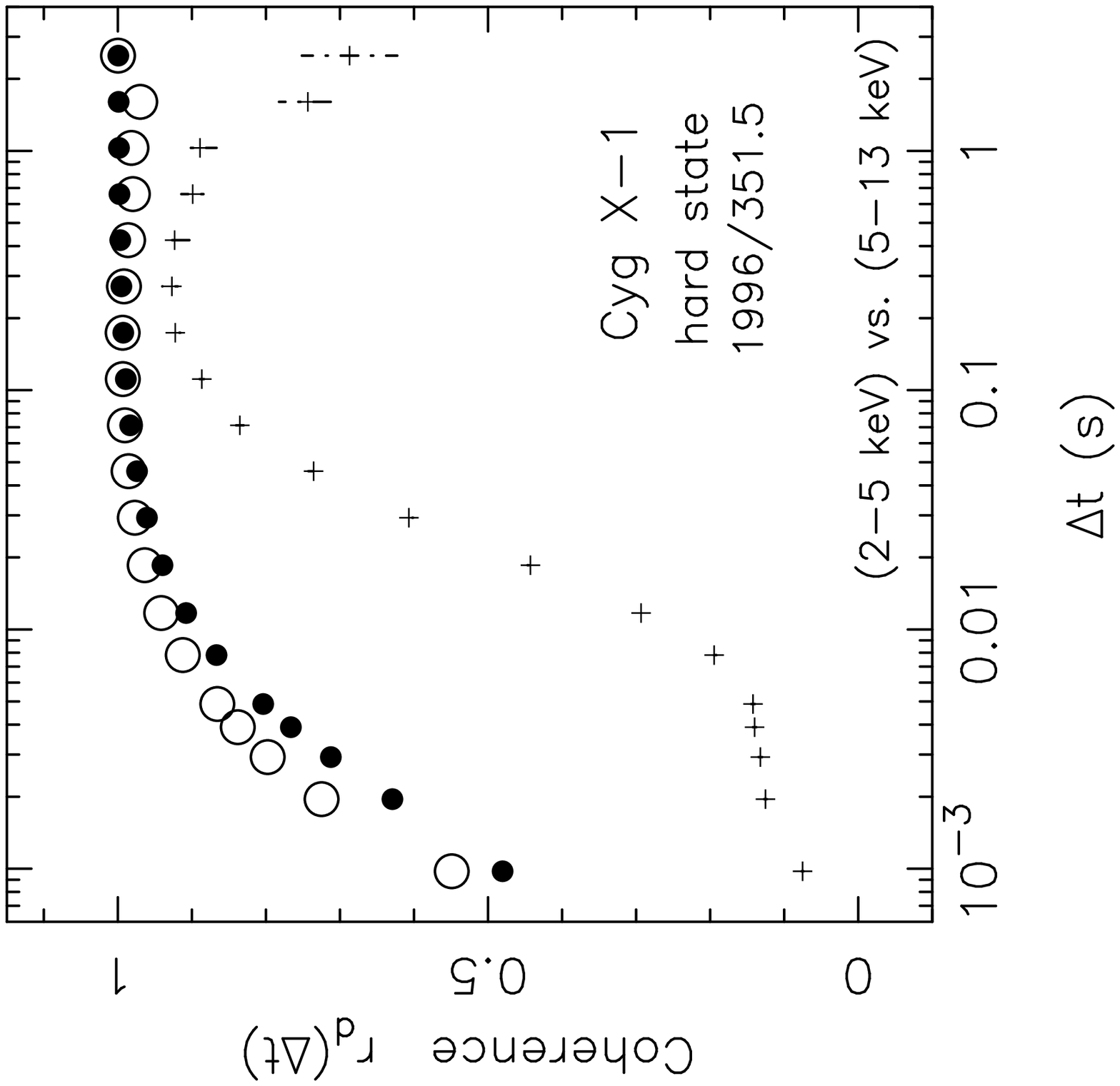}{20pt}{-90}{40}{25}{-50}{152}
\vspace{-4mm}
\caption{Variability coherence of different component vs. time scale of Cyg X-1. 
  {\it Filled circle}: total light curve. {\sl Plus}: shot
component. {\it Circle}: steady component.  
\label{fig9}}
\end{figure}

\begin{figure}
\vspace{3.5cm}
\epsscale{1.0}
\plotfiddle{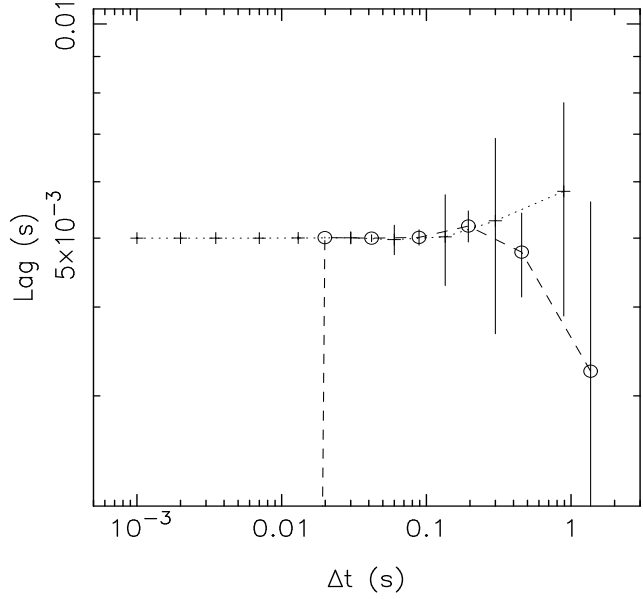}{20pt}{-90}{50}{45}{-180}{178}
\vspace{2.5cm}
\caption{Time lag vs. time scale of two light curves of white noise with 5 ms time lag. 
{\sl Circle and dashed line}:
calculated by Fourier cross spectrum. {\sl Plus and dotted line}: by modified cross-correlation
function in the time domain.
\label{fig10}}
\end{figure}

 \begin{figure}
\vspace{5cm}
\epsscale{1.0}
\plotfiddle{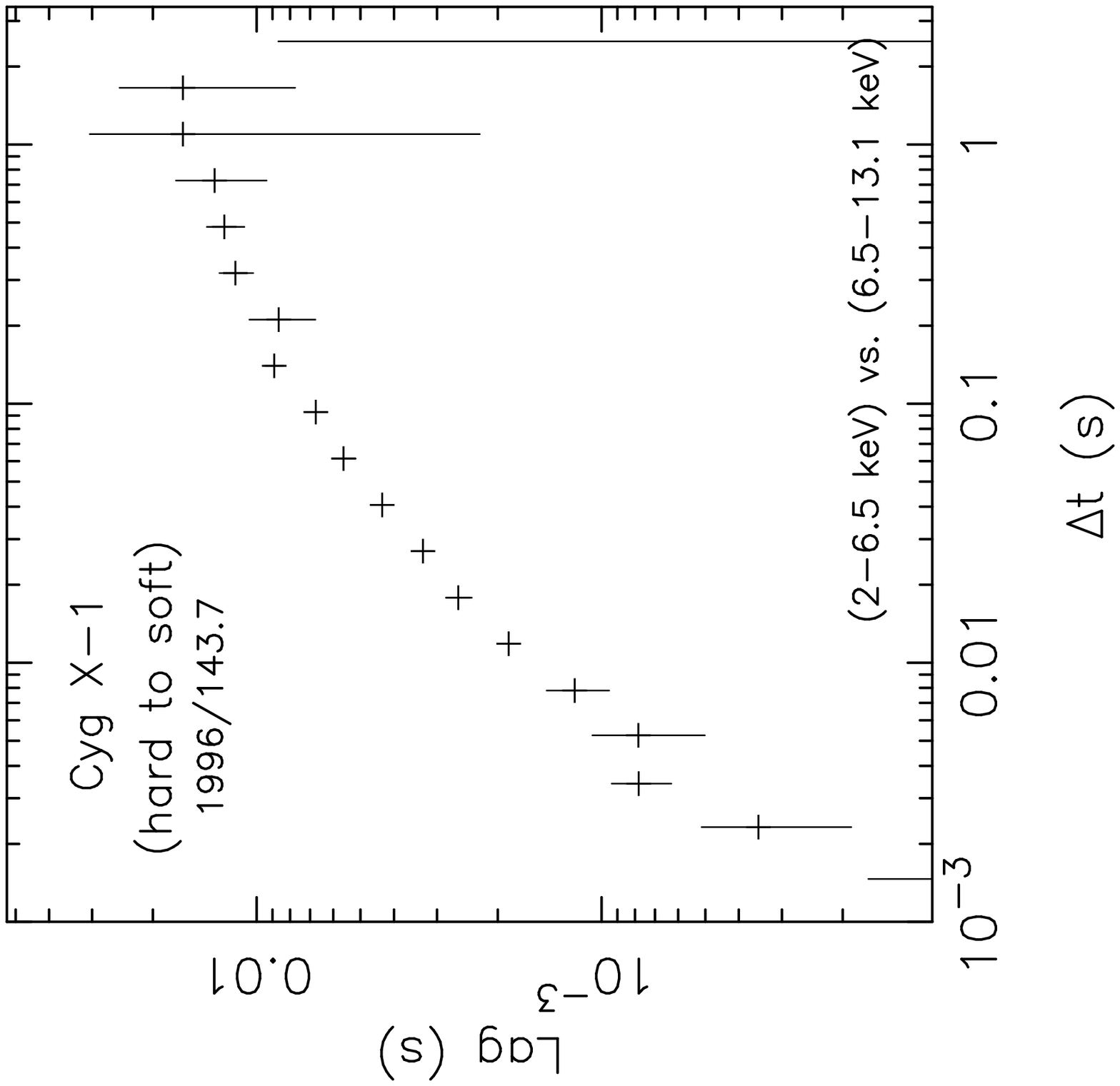}{20pt}{-90}{40}{25}{-260}{178}
\plotfiddle{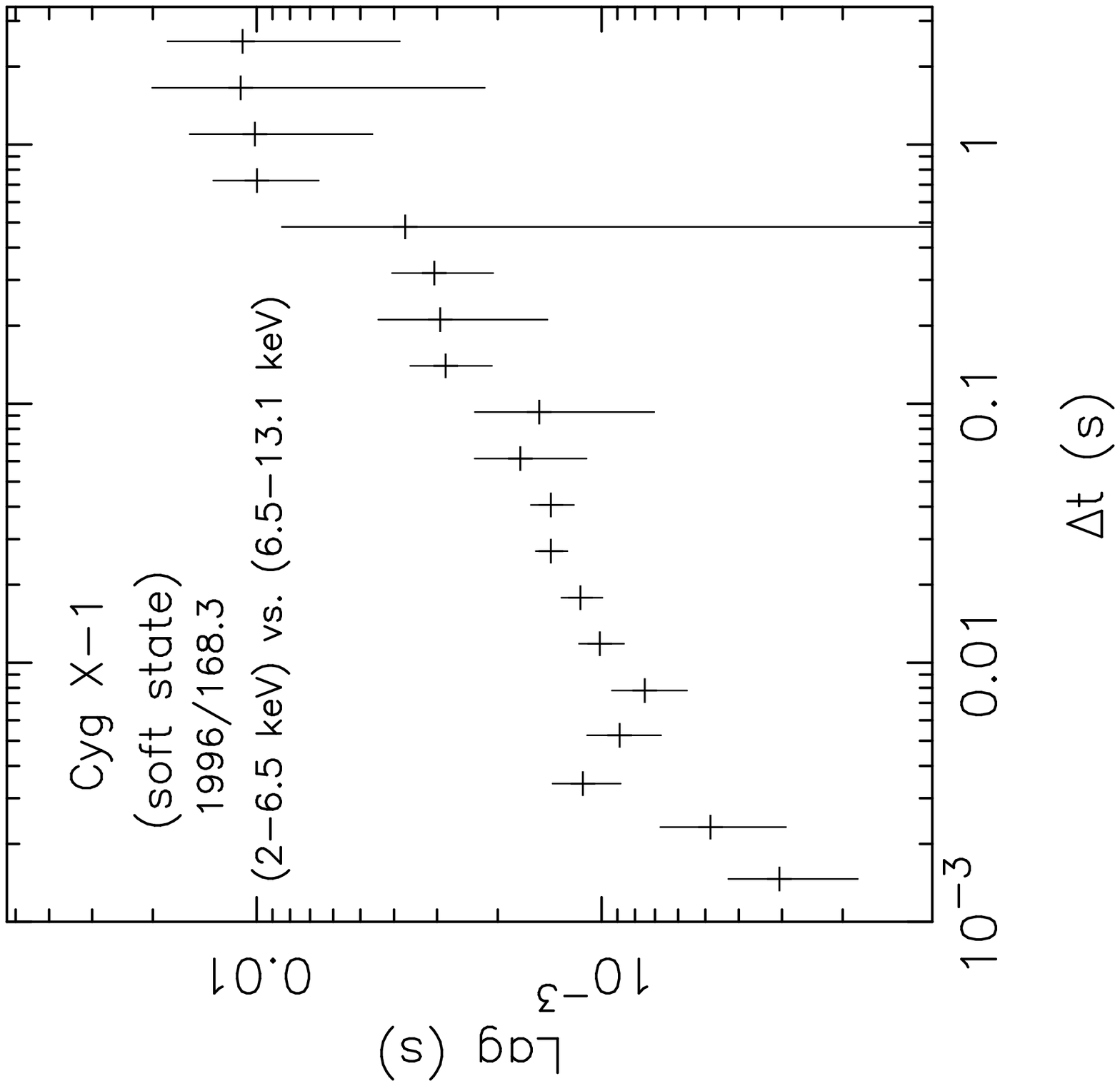}{20pt}{-90}{40}{25}{-50}{212}
\plotfiddle{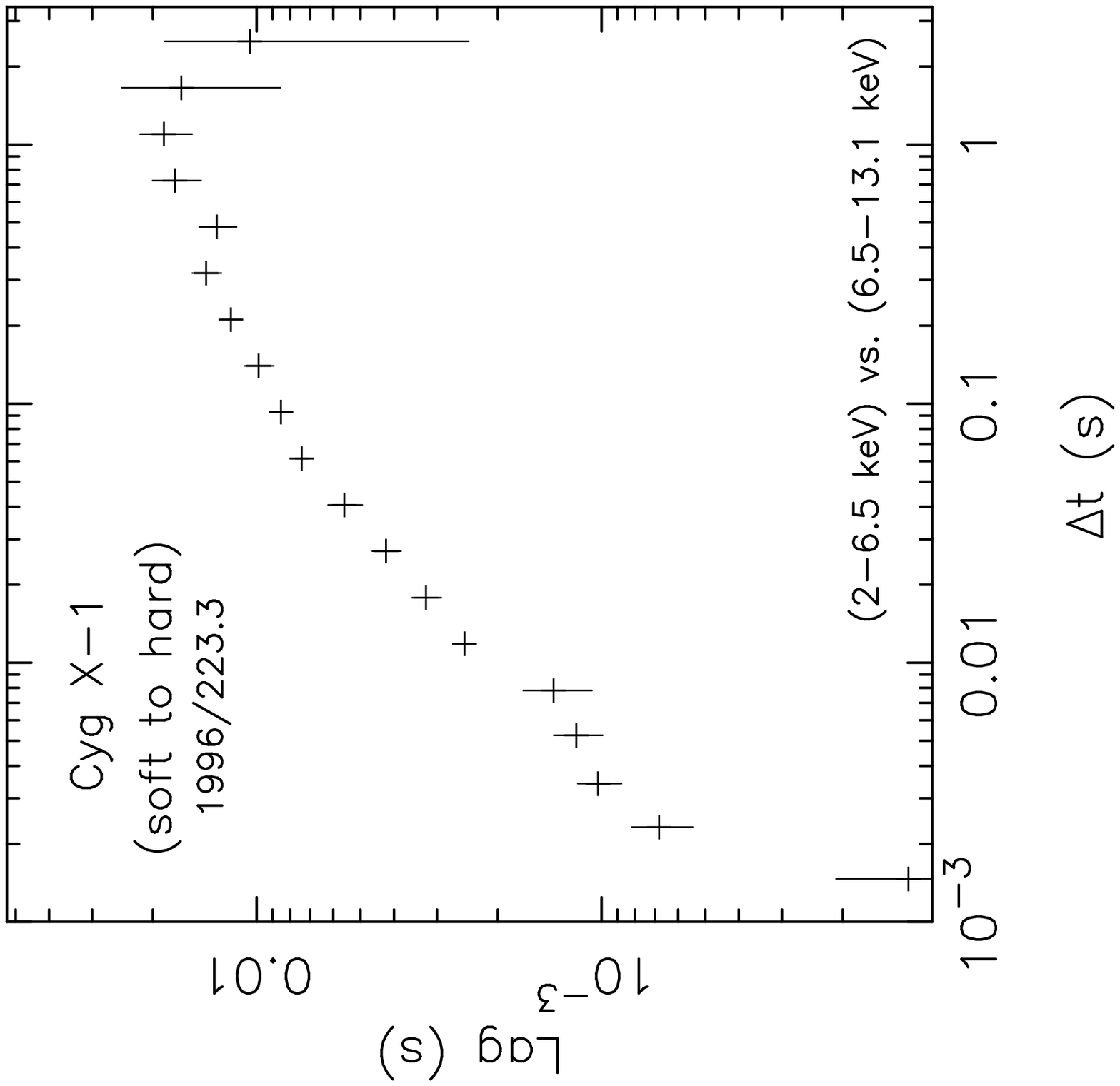}{20pt}{-90}{40}{25}{-260}{119}
\plotfiddle{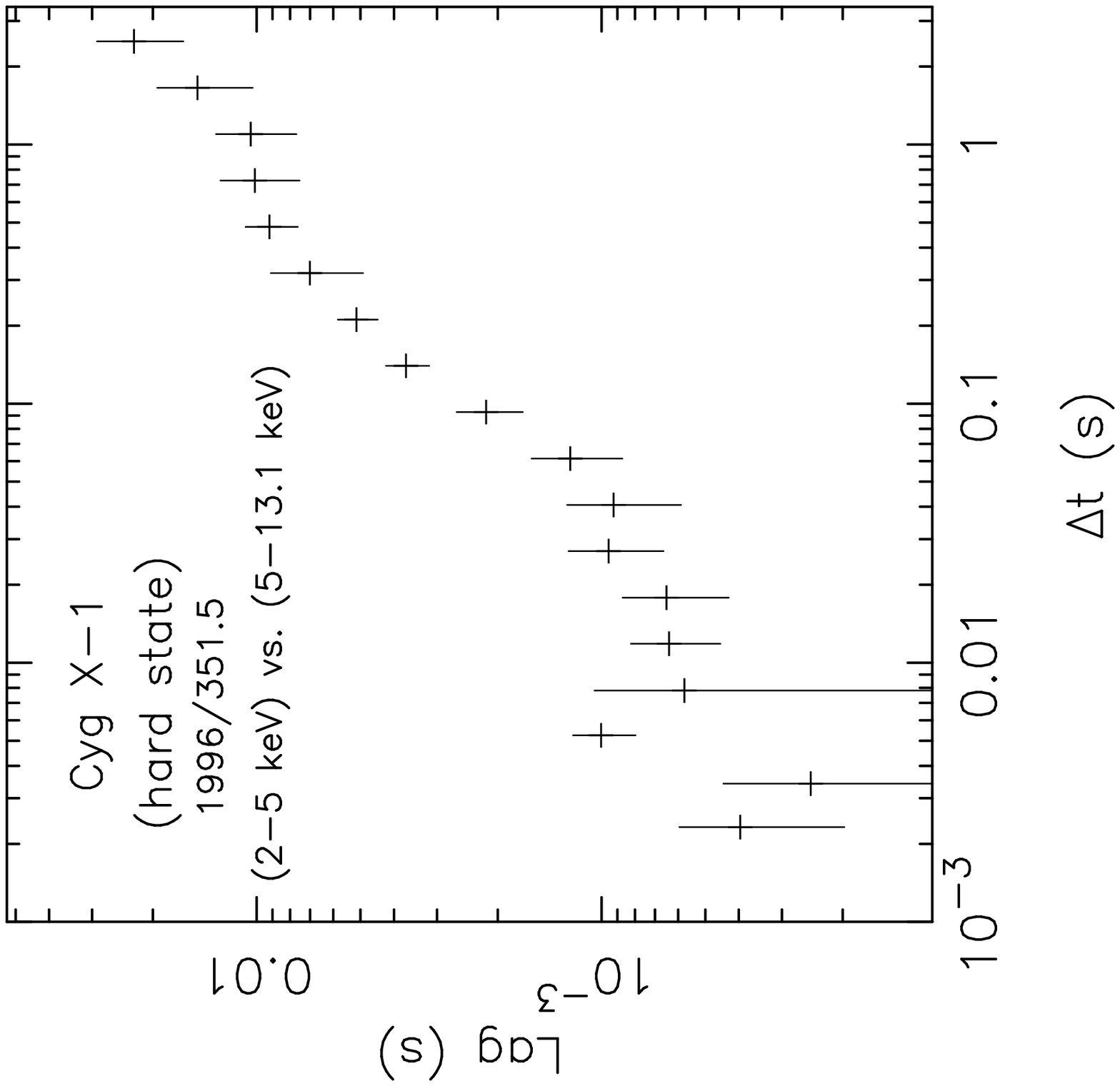}{20pt}{-90}{40}{25}{-50}{152}
\vspace{-4mm}
\caption{Hard X-ray lag vs. time scale of Cyg X-1 \label{fig11}}
\end{figure}

\begin{figure}
\vspace{4.5cm}
\epsscale{1.0}
\plotfiddle{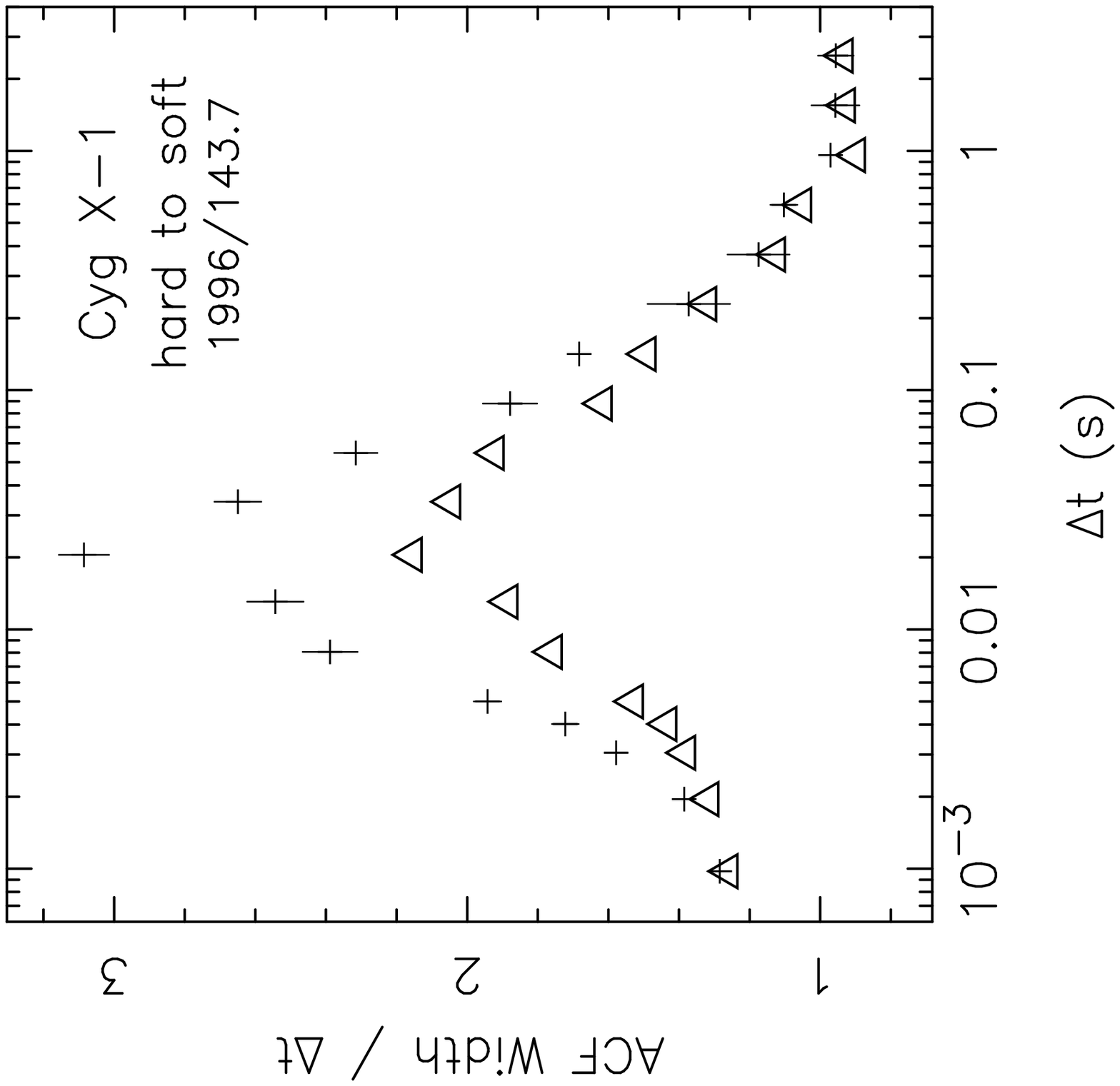}{20pt}{-90}{40}{25}{-260}{178}
\plotfiddle{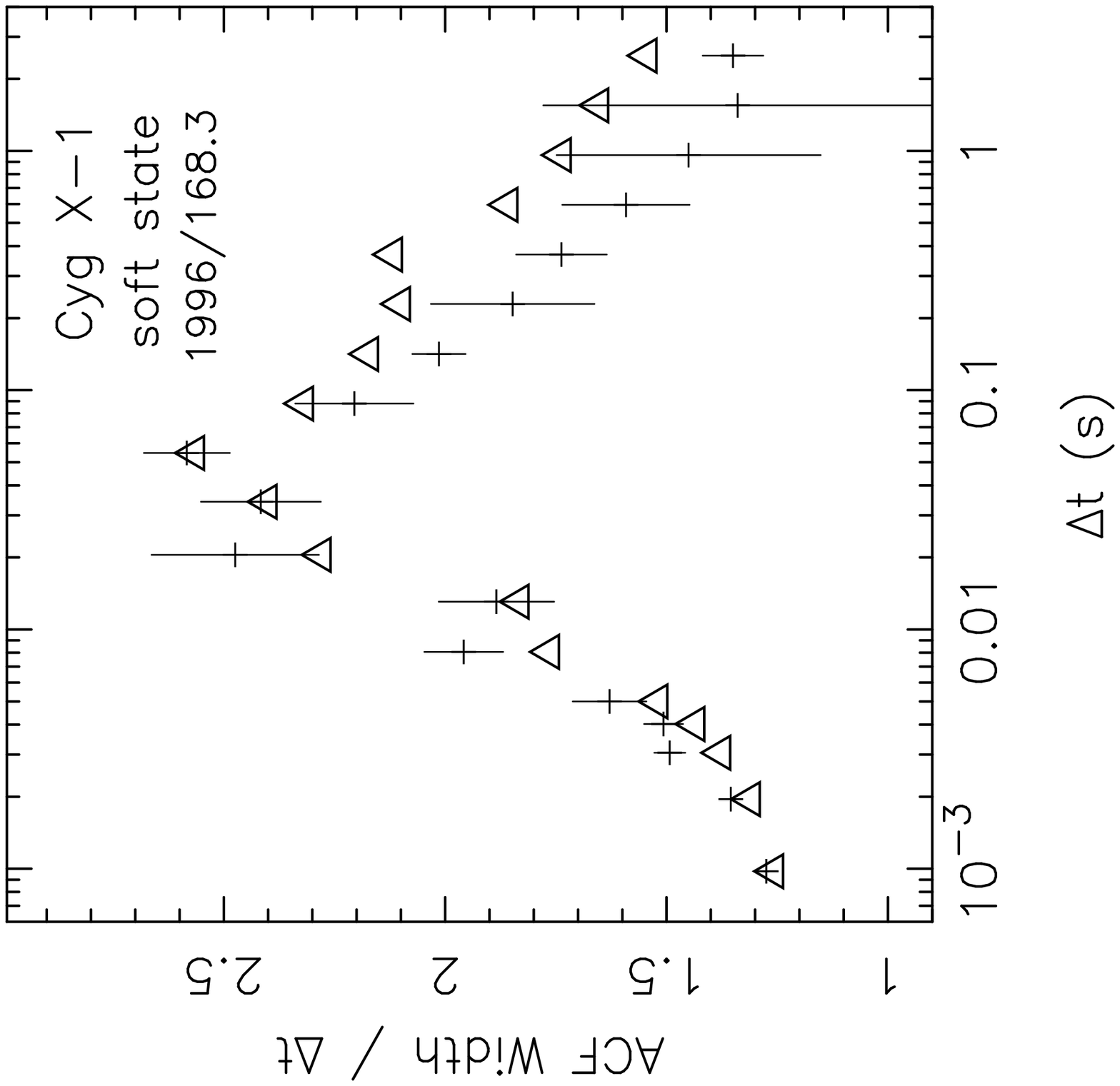}{20pt}{-90}{40}{25}{-50}{212}
\plotfiddle{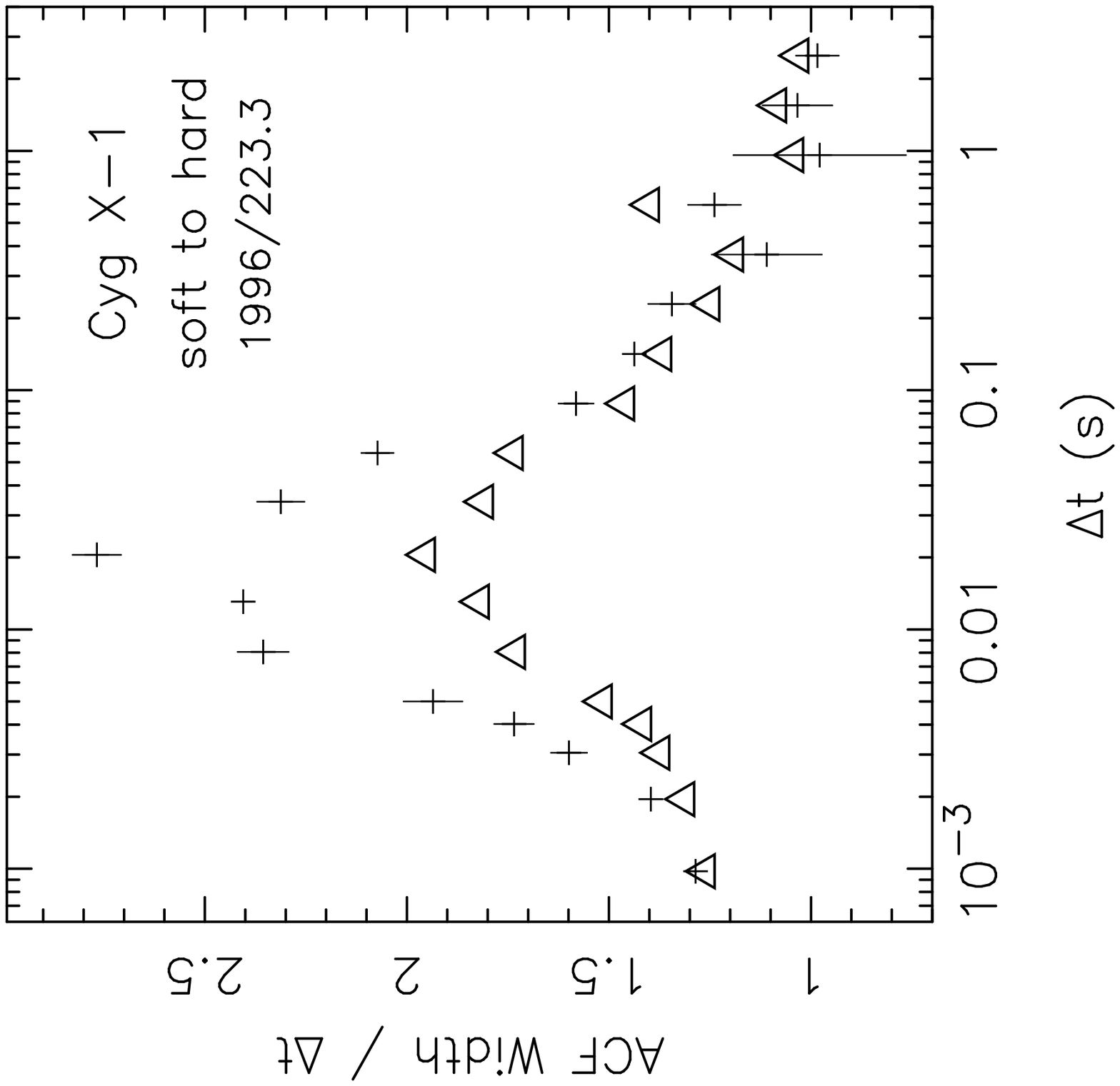}{20pt}{-90}{40}{25}{-260}{118}
\plotfiddle{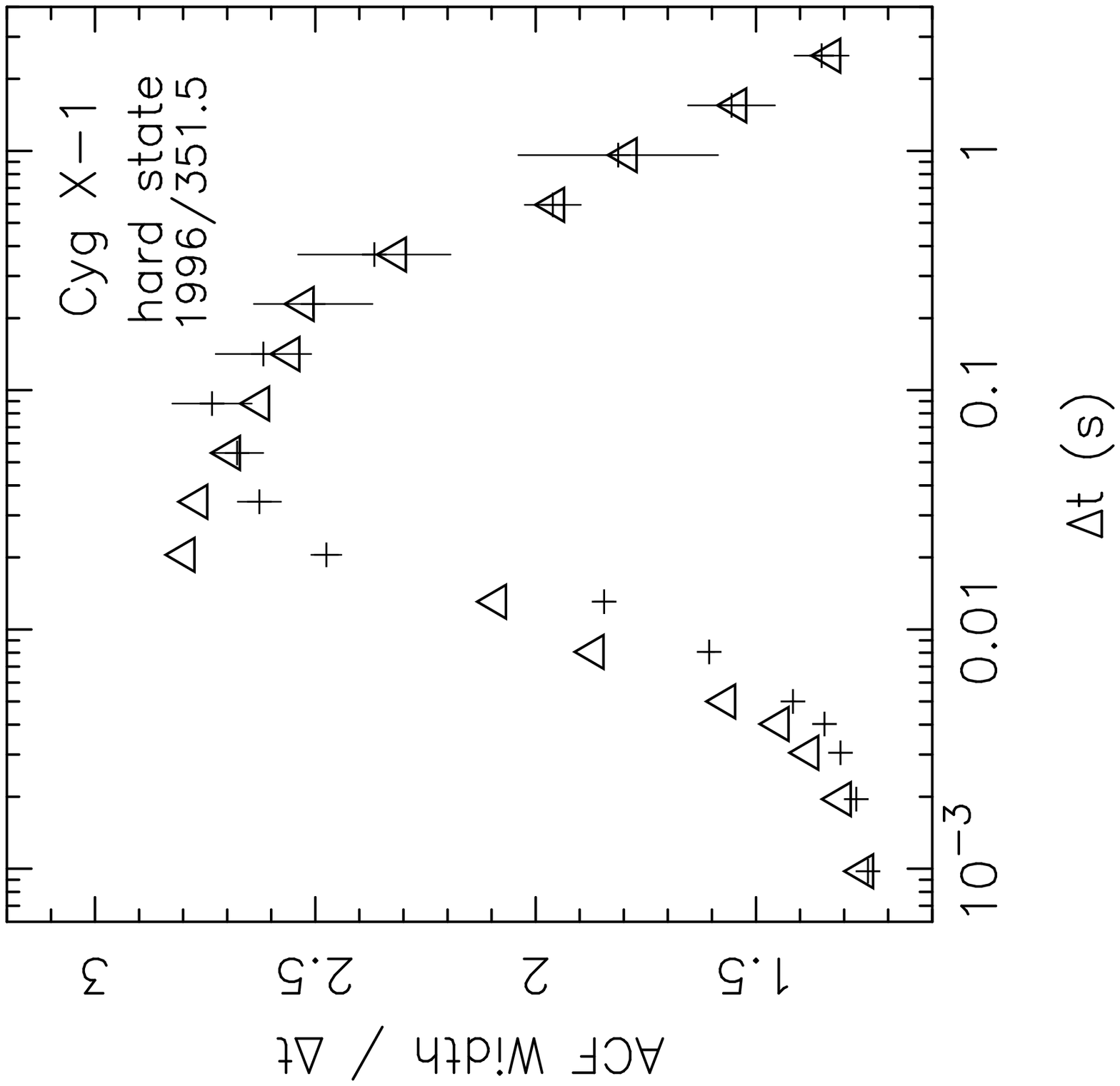}{20pt}{-90}{40}{25}{-50}{152}
\vspace{-4mm}
\caption{Variation  duration (MACF width) vs. time scale of Cyg X-1. {\it Plus}: low-energy band (2-5 keV for hard state, 2-6.5 keV
for other states). {\it Triangle}: high-energy band (5-13 keV for hard state, 
6.5-13 keV for other states)
\label{fig12}}

\end{figure}
 \begin{figure}
\vspace{4.3cm}
\epsscale{1.0}
\plotfiddle{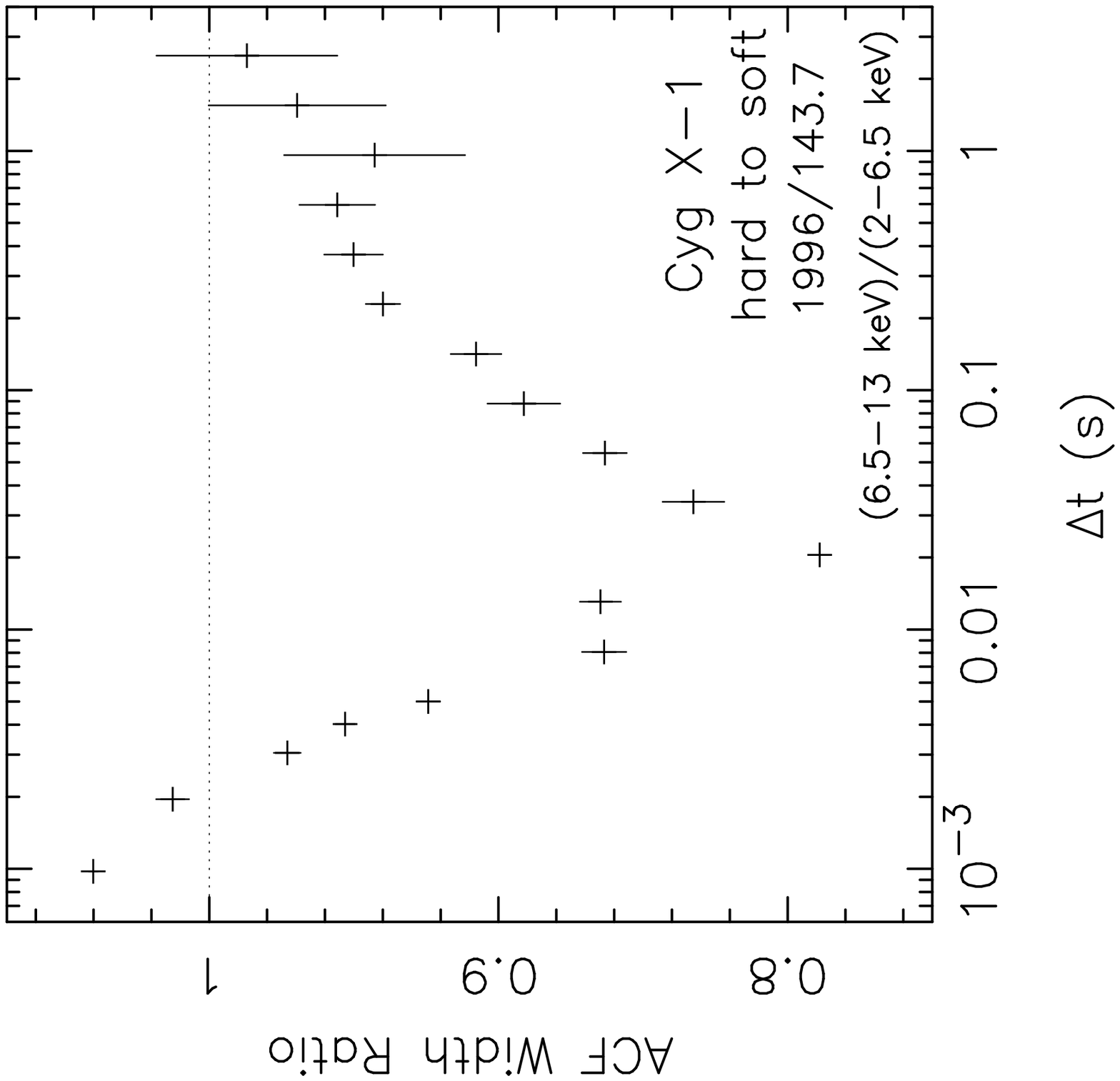}{20pt}{-90}{40}{25}{-260}{178}
\plotfiddle{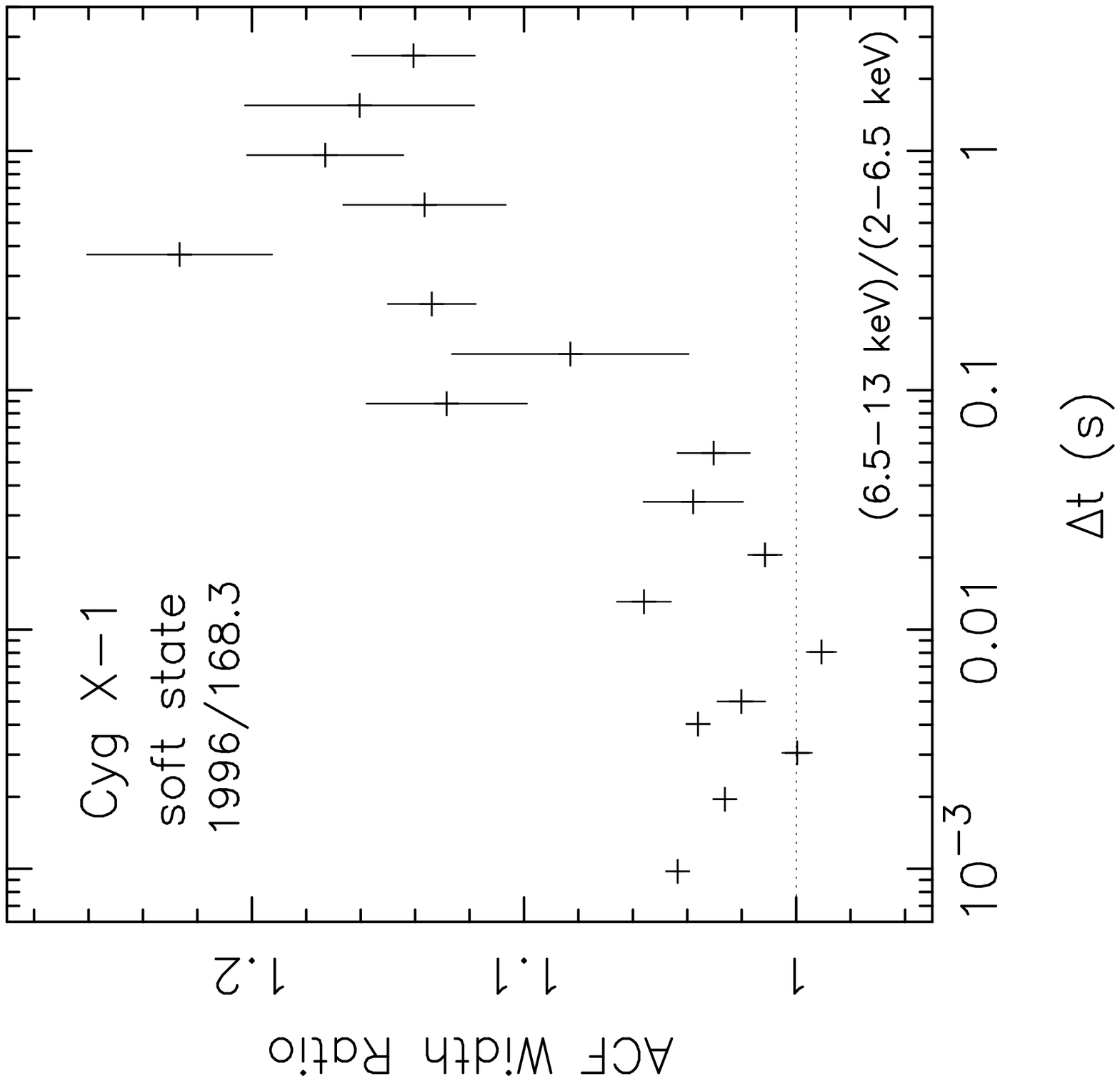}{20pt}{-90}{40}{25}{-50}{212}
\plotfiddle{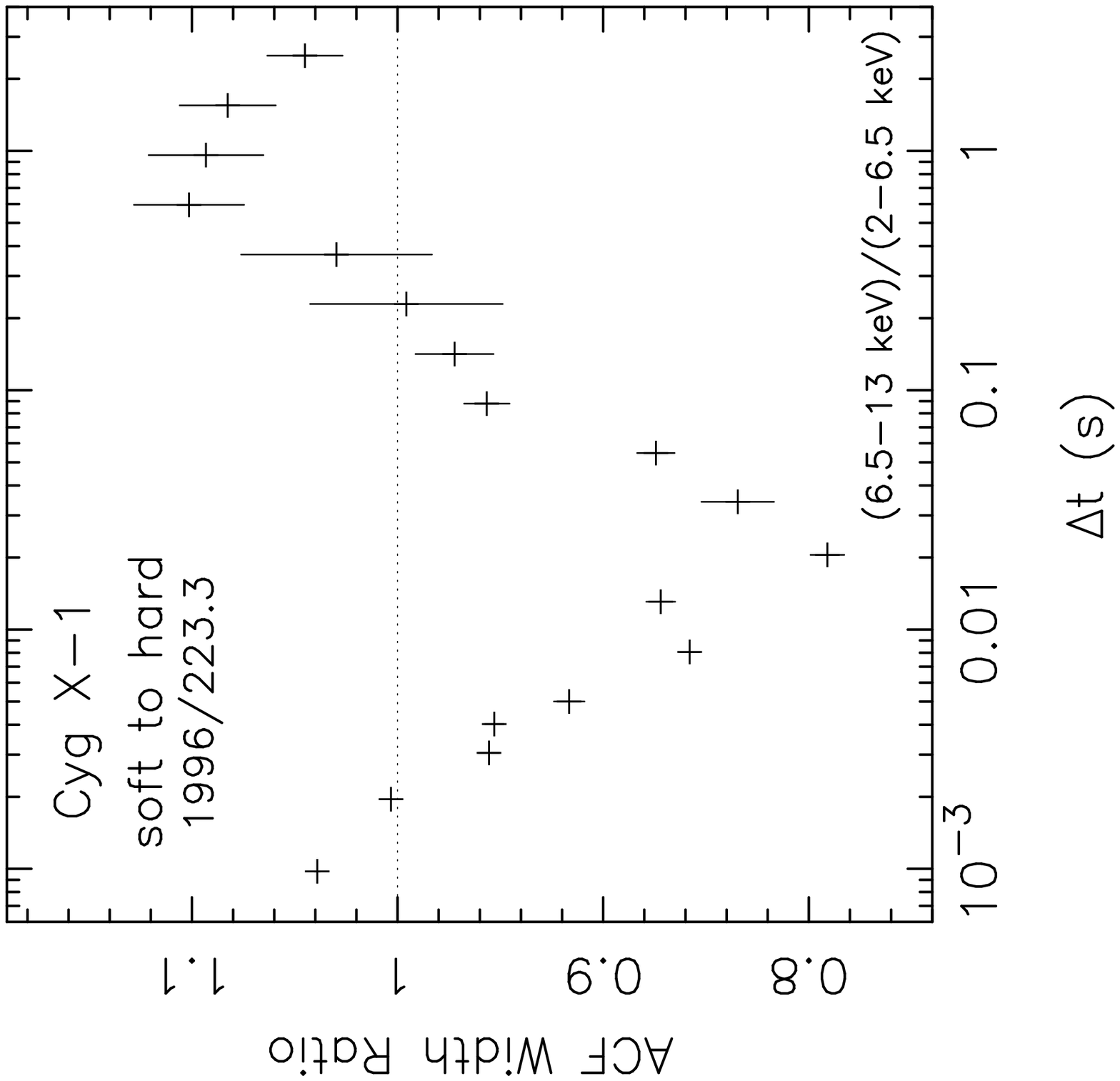}{20pt}{-90}{40}{25}{-260}{118}
\plotfiddle{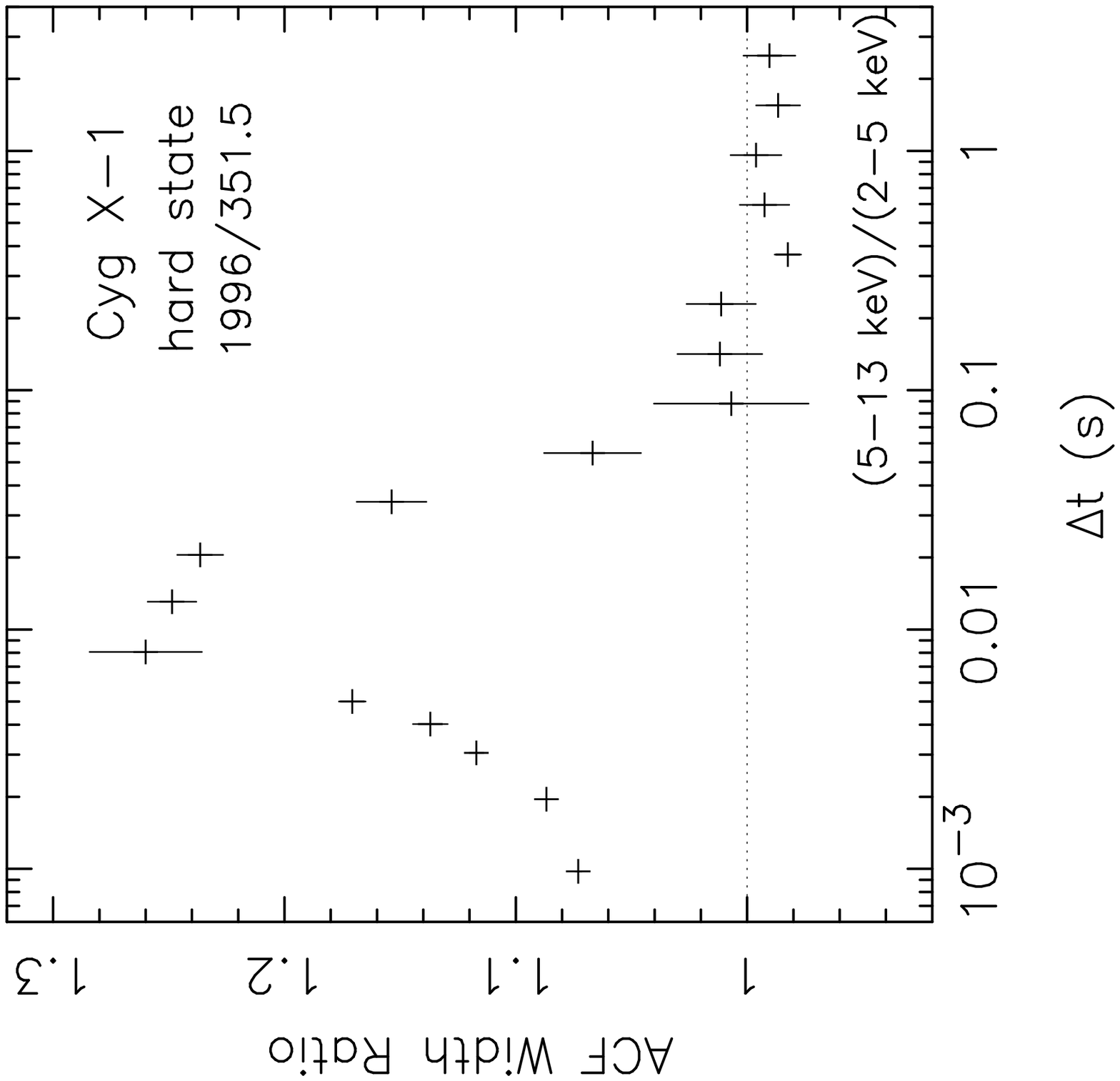}{20pt}{-90}{40}{25}{-50}{152}
\vspace{-4mm}
\caption{MACF width ratio vs. time scale of Cyg X-1 \label{fig13}}
\end{figure}

 \begin{figure}
\vspace{4.3cm}
\epsscale{1.0}
\plotfiddle{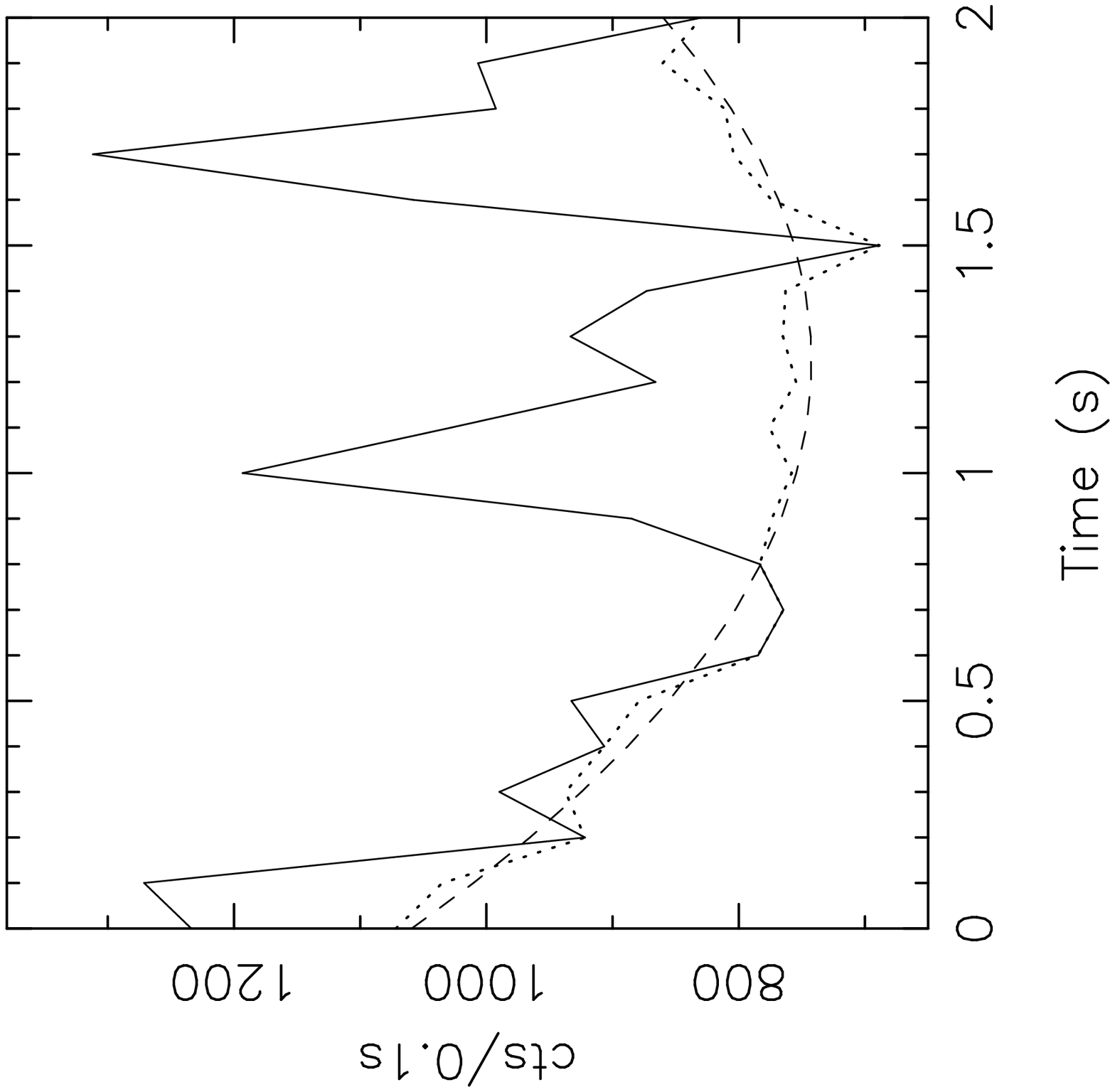}{20pt}{-90}{40}{25}{-260}{178}
\plotfiddle{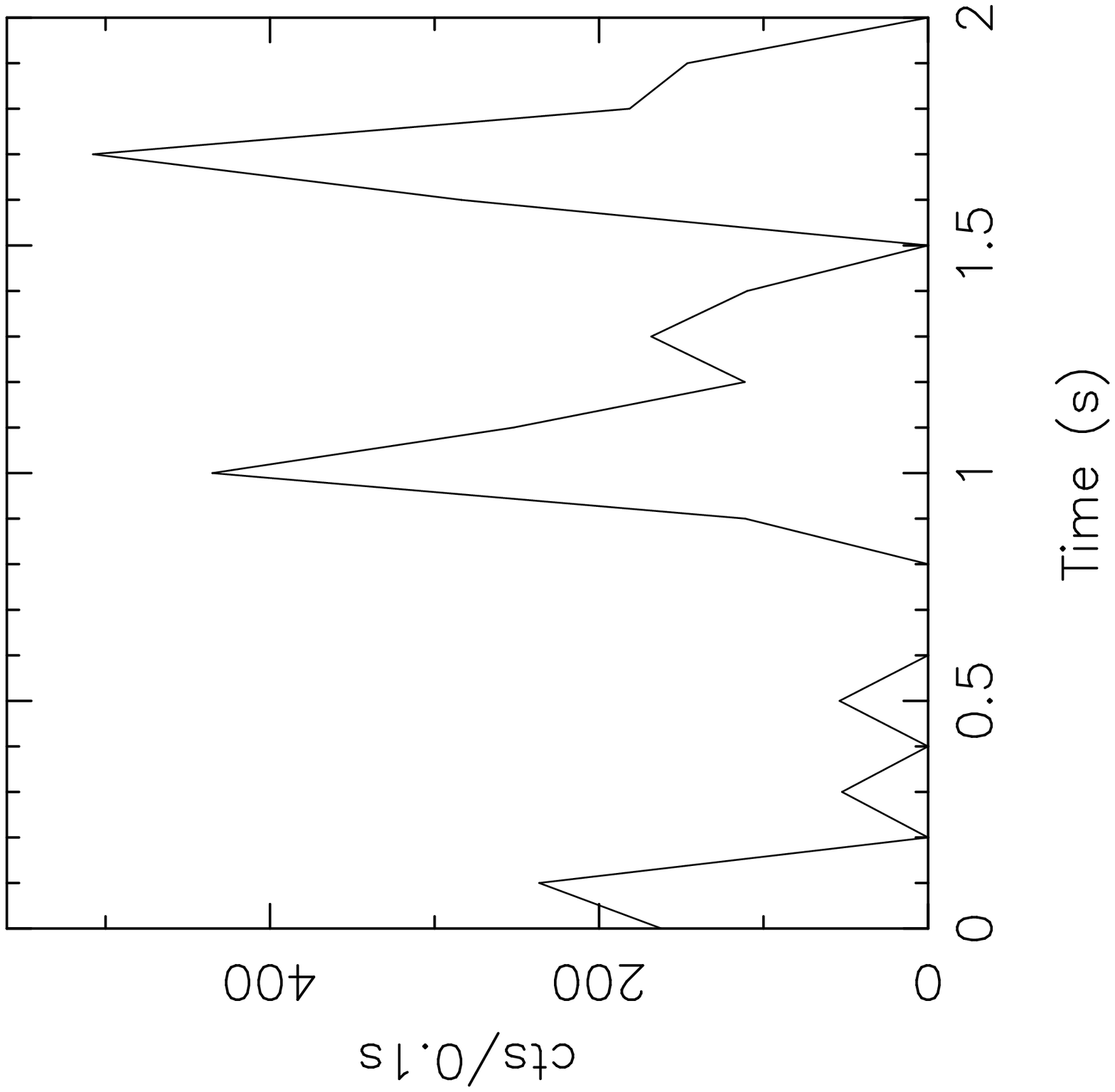}{20pt}{-90}{40}{25}{-50}{212}
\vspace{-26mm}
\caption{Different components of light curve. {\sl Left}: {\it solid line} - total 
light curve; {\it dotted line} - steady component;  {\it dashed line} - least-square 
polynomial of steady component. {\sl Right}: shot component 
  \label{fig14}}  
\end{figure}

\end{document}